\documentclass[prb,floatfix,groupedaddress,amsmath,amssymb,notitlepage]{revtex4-1}

\usepackage{graphicx}
\usepackage{dcolumn}
\usepackage{bm}
\usepackage[table]{xcolor}
\usepackage{units}

\usepackage{natbib}

\usepackage{subfigure}

\newcommand\mcn[2]{\multicolumn{#1}{c}{#2}}

\usepackage{varioref}
\usepackage[colorlinks]{hyperref}

\providecommand\element[1]{\mbox{#1}\mbox{\kern1pt}}

\newcolumntype{C}{>{$}c<{$}}
\newcolumntype{L}{>{$}l<{$}}
\newcolumntype{F}{>{\hphantom{mist}}c}
\newcolumntype{H}[1]{>{\centering\arraybackslash}m{#1}}

\definecolor{hellorange}{HTML}{FF9900}
\definecolor{gelbbraun}{HTML}{FFCC99}

\begin{document}

\title{Another Survey of Foundational Attitudes Towards Quantum Mechanics}

\author{Christoph Sommer}
\email{c.sommer@uni-muenster.de}
\affiliation{Westf\"alische Wilhelms-Universit\"at M\"unster, 
 D-48149 M\"unster, Germany}

\begin{abstract}

Although it has been almost 100 years since the beginnings of quantum mechanics, the discussions about its interpretation still do not cease. Therefore, a survey of opinions regarding this matter is of particular interest.
This poll was conducted following an idea and using the methodology of Schlosshauer et al.\cite{Zeilinger}, but among a slightly different group. It is supposed to give another snapshot of attitudes towards the interpretation of quantum mechanics and keep discourse about this topic alive.

\end{abstract}

\maketitle

\section{\label{sec:level1}Introduction}
The following survey is by no means the first one ever conducted on this topic. For instance in 1997, some 48 participants of a UMBC quantum mechanics workshop were asked by Tegmark \cite{Tegmark} about their favored interpretation of quantum mechanics. Although the Copenhagen interpretation won obtaining 13 votes (only topped by the "None of the above/undecided" category that received 18 votes), the many-worlds interpretation followed closely with 8 votes. According to Tegmark, this "indicated a rather striking shift in opinion compared to the old days when the Copenhagen interpretation reigned supreme."\\
Most recently, Schlosshauer et al.\cite{Zeilinger} published a poll conducted among the 35 participants of the conference "Quantum Physics and the Nature of Reality" at the International Academy Traunkirchen, Austria, organized by A.~Zeilinger.
This latter study serves as a model for this work. Questionnaire and methodology are chosen equal with one exception: To the question "What is your favorite interpretation of quantum mechanics" (question 12), the answer "shut up and calculate" has actually been included as already suggested by Schlosshauer et al. The questioned group is similar in affiliations to theirs, but probably at an earlier stage of their careers. It will be interesting to compare the results and give a potentially different snapshot of opinions towards the foundations of quantum mechanics.

The occasion for this survey was given in early 2013 by a conference about the "Philosophy of Quantum Mechanics" that took place in a remote hut in the black forest, Germany.
Of the 21 participants of this conference, 18 turned in their completed questionnaires. Of these, 12 stated their affiliation as physics, 5 as mathematics and 2 as philosophy (multiple or no answers were possible). Most of them are late master students or early PhD students.
Naturally, such a small sample is not representative for all physicists or people concerned with quantum mechanics. Hence, this study is of rather informative character, but may be nonetheless interesting. The questionnaire and methodology have been chosen intentionally almost equal to that of Schlosshauer et al. \cite{Zeilinger} in order to allow comparison, not taking the small size of the sample all too serious.

This report is organized as follows. In Sec. \ref{sec:results} the results of the poll are discussed.
In the subsequent Sec. \ref{sec:corr}, correlations among the answers are shown.
Discussion and summary of the results are given in Sec. \ref{sec:dis} and \ref{sec:sum}.
For the sake of completeness, in Appendix \ref{sec:corrtab} the whole correlation table is added.

\section{\label{sec:results}Results}

In this section, the results of the poll are to be discussed. The questionnaires were handed out at the end of the conference and collected soon afterwards. Multiple answers were possible and sometimes no answer was checked. No additional explanation concerning the meaning of questions or answers was given.

\begin{center}
\includegraphics[width=.7\linewidth]{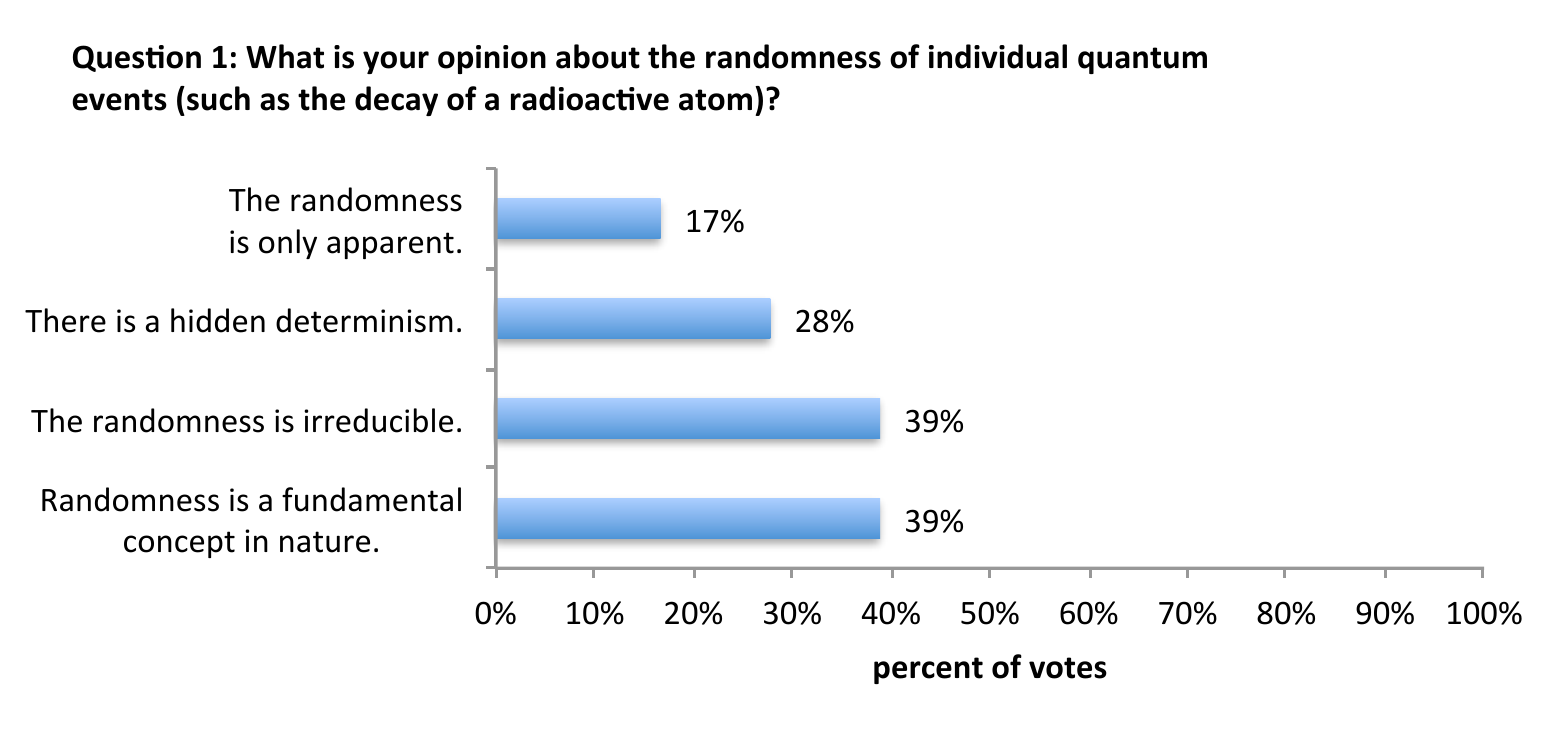}
\end{center}

Quite contrary to the polls of Schlosshauer et al.\cite{Zeilinger} and Tegmark \cite{Tegmark}, 3 people favored the de Broglie-Bohm \cite{deBroglie,Bohm1,Bohm2} and no one the Everett interpretation (see Ref. \cite{Tegmark} and references therein) of quantum mechanics (compare question 12). The higher support for the hidden determinism can thus be understood, because the interpretation of Bohm might have been considered well compatible with this answer by some participants.

\begin{center}
\includegraphics[width=.7\linewidth]{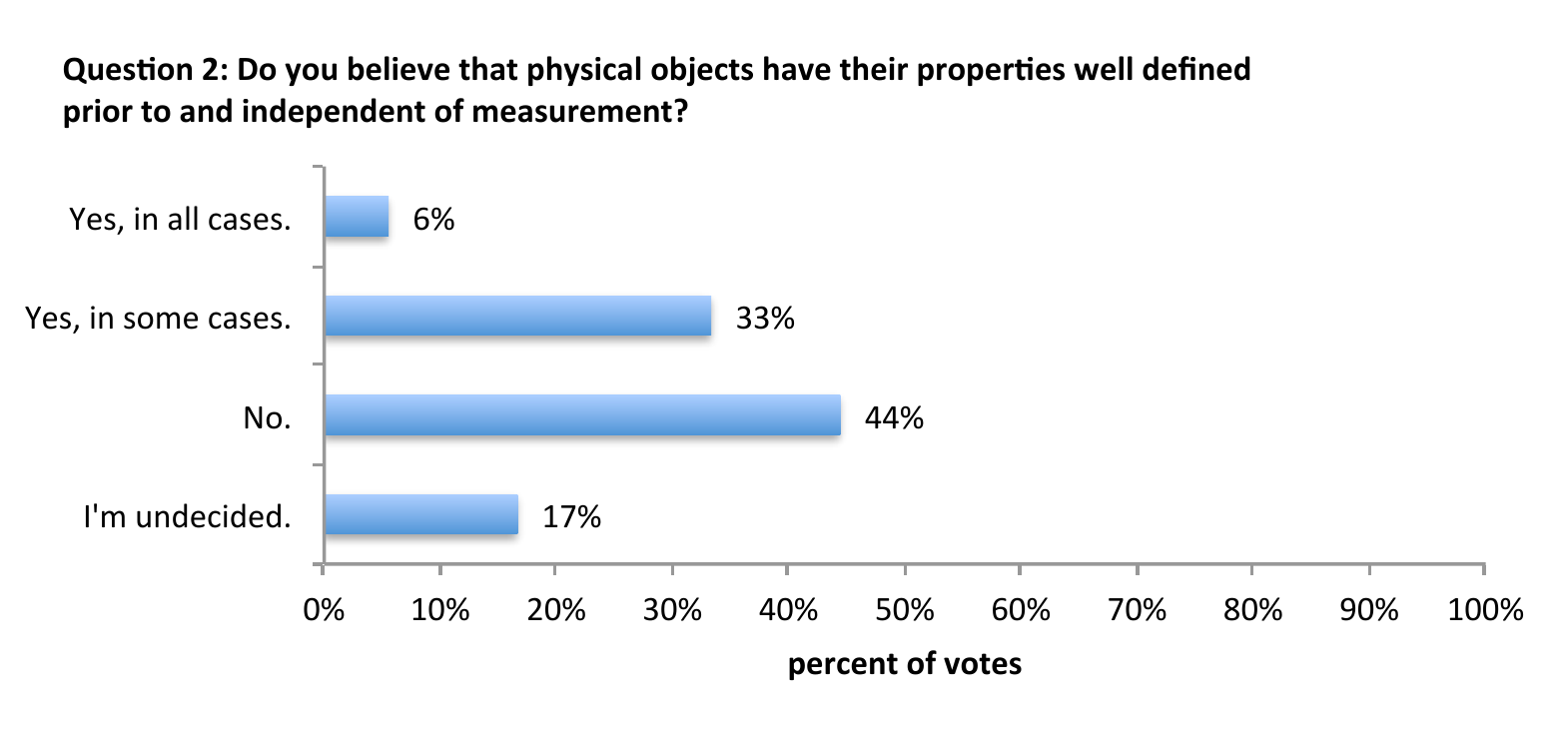}
\end{center}

This question was answered by trend like the corresponding one of Schlosshauer et al\cite{Zeilinger}. There is a virtual draw between affirmative and negative answers. As could be understood from some write-ins, the vagueness in the words "properties" and "measurement" added to this draw.

\begin{center}
\includegraphics[width=.7\linewidth]{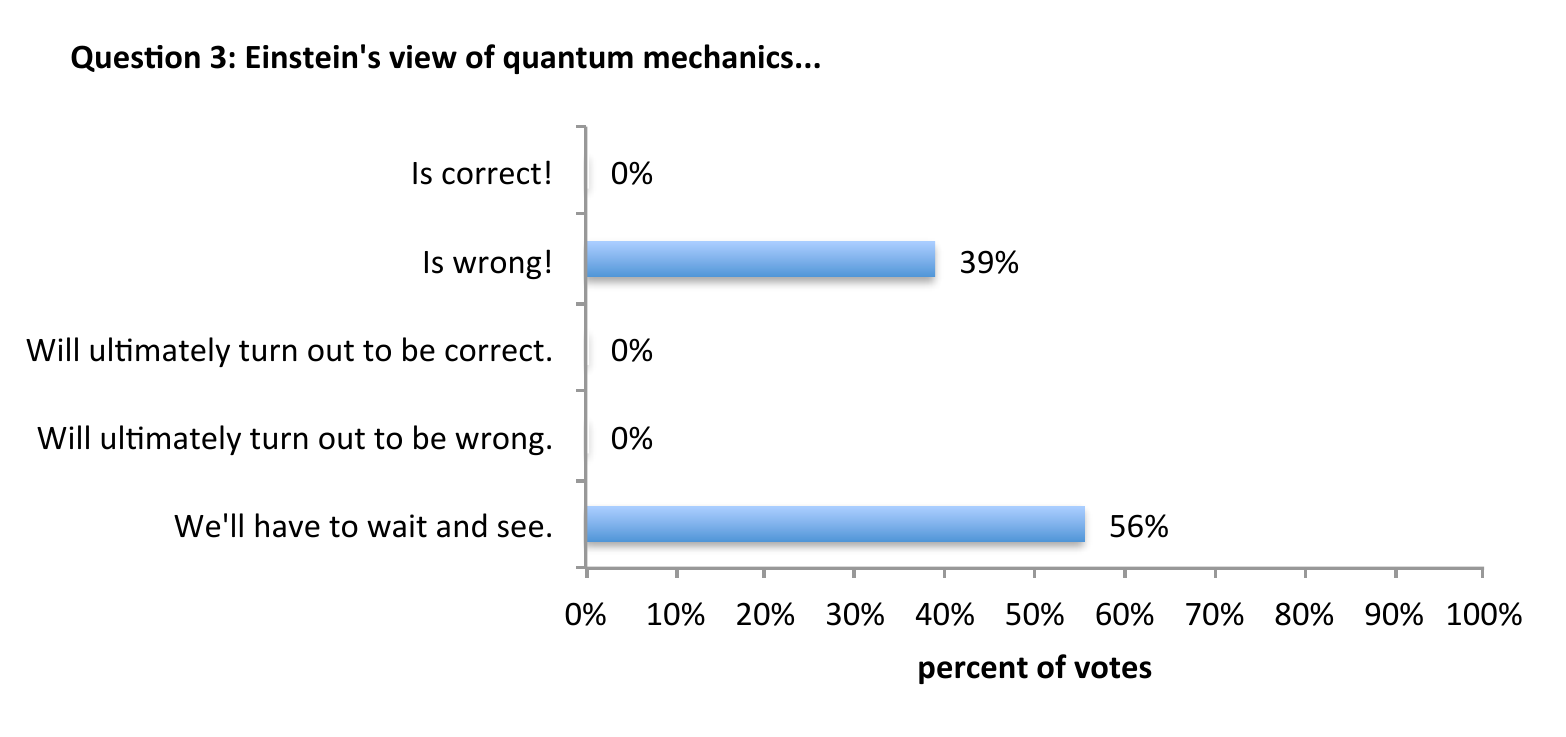}
\end{center}

The result is clearly negative here about Einstein's view of quantum mechanics. Although in the first question, 17\% asserted that the randomness of individual quantum events is only apparent, all participants either found Einstein's view of quantum mechanics wrong or at least remained hesitant. Important for this outcome is also the fact that it was never concretized what Einstein's view actually is.

\begin{center}
\includegraphics[width=.7\linewidth]{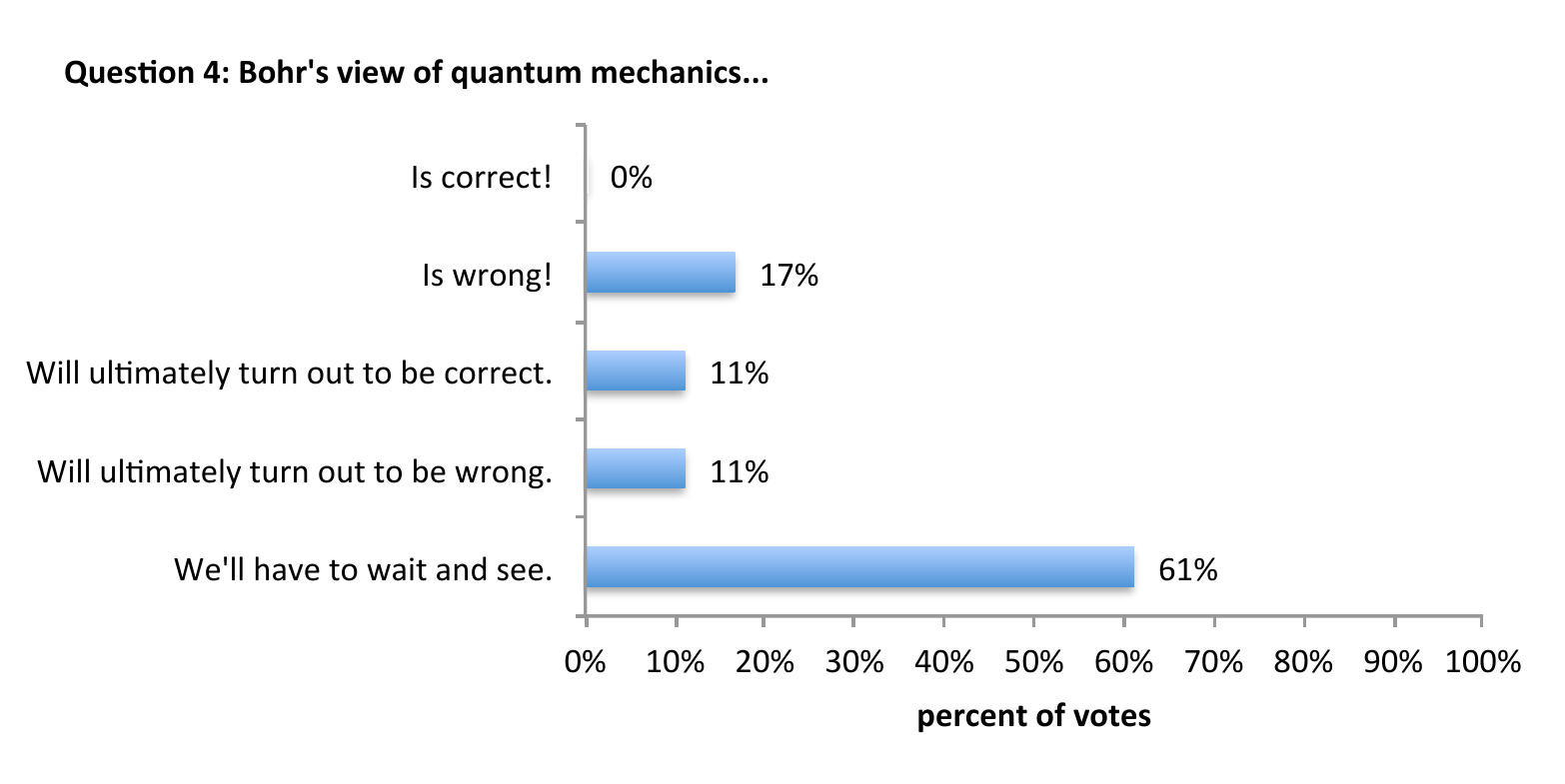}
\end{center}

In what concerns Bohr's view of quantum mechanics, the great majority of people maintain a rather reserved attitude and only few were willing to support or decline Bohr's view. However, nobody felt confident to call Bohr's view correct per se.

\begin{center}
\includegraphics[width=.7\linewidth]{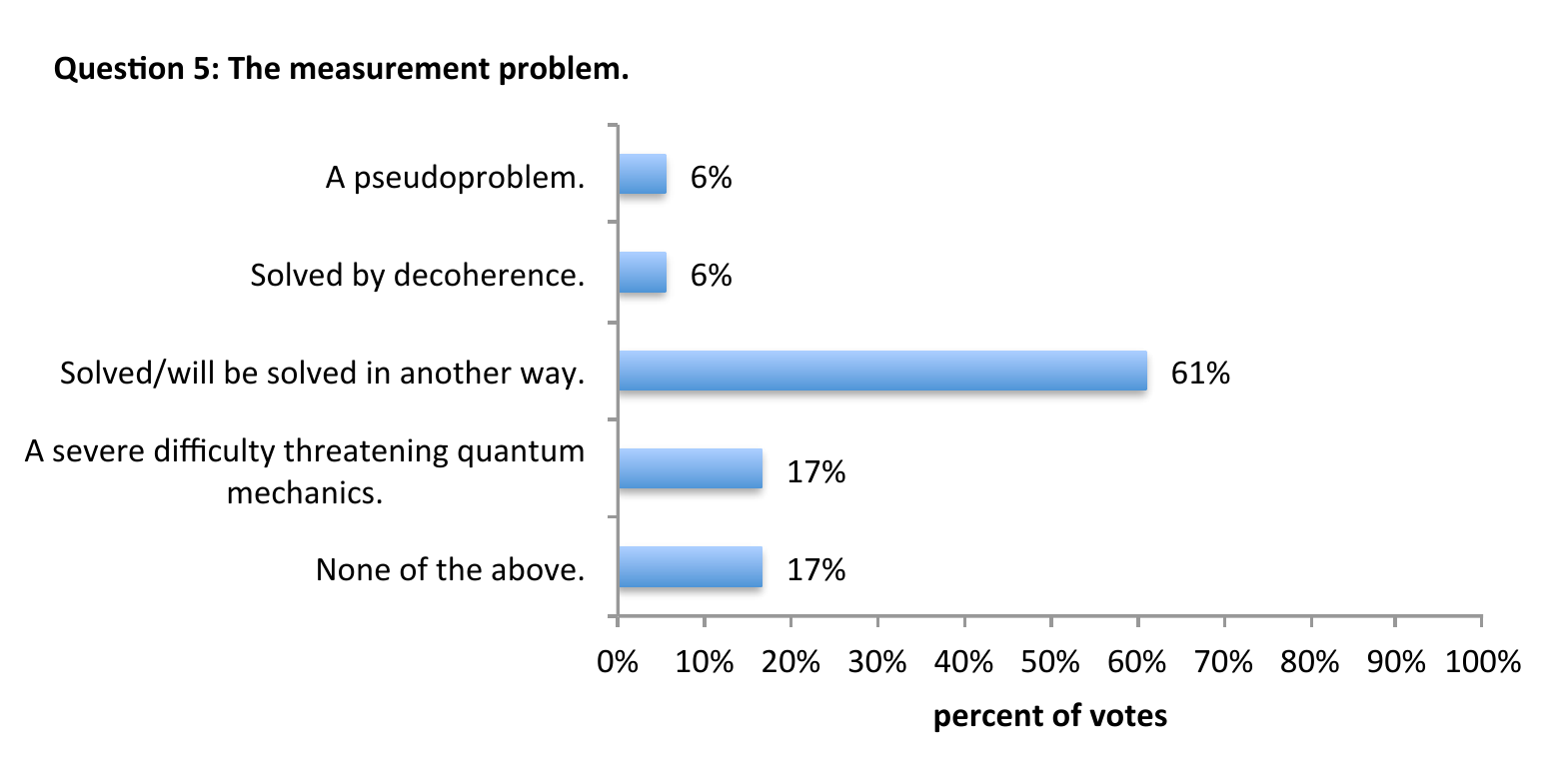}
\end{center}

Obviously, most participants regard the measurement problem as solved or at least not to pose such a big threat for the general theory. On some questionnaires, though, the words "will be solved in another way" were underlined or highlighted in another way, making clear that this problem may be still all but settled.

\begin{center}
\includegraphics[width=.7\linewidth]{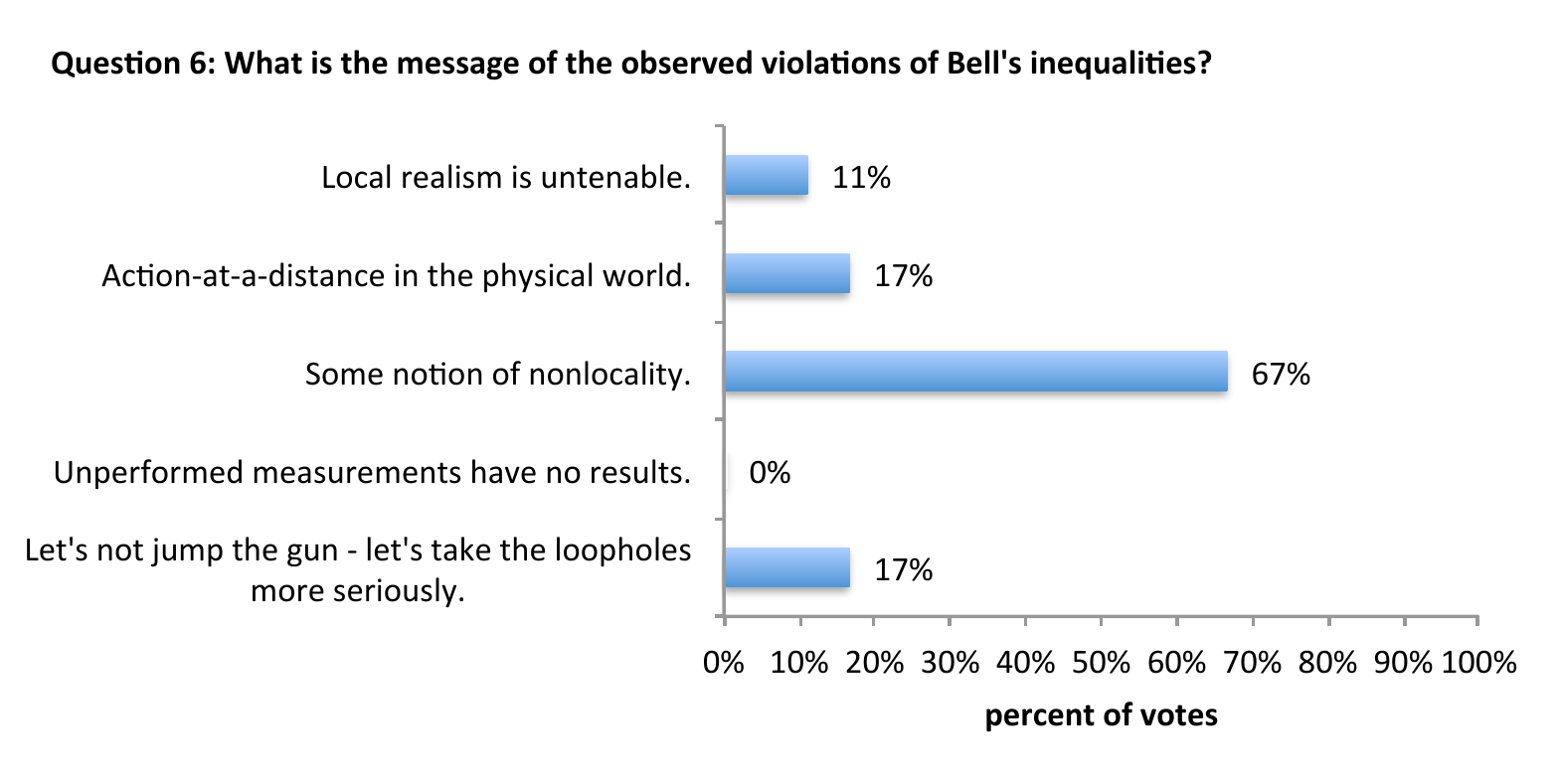}
\end{center}

Most participants found that the implication of Bell's inequalities is "some notion of nonlocality". Since the word "realism" always led to lively discussions at the conference, the small support for the first answer can be understood, since many people may have wanted to evade the complex of problems that is connected to this notion.
The participants unanimously agreed upon the fact that the implication is \textit{not} that unperformed measurements have no results.

\begin{center}
\includegraphics[width=.7\linewidth]{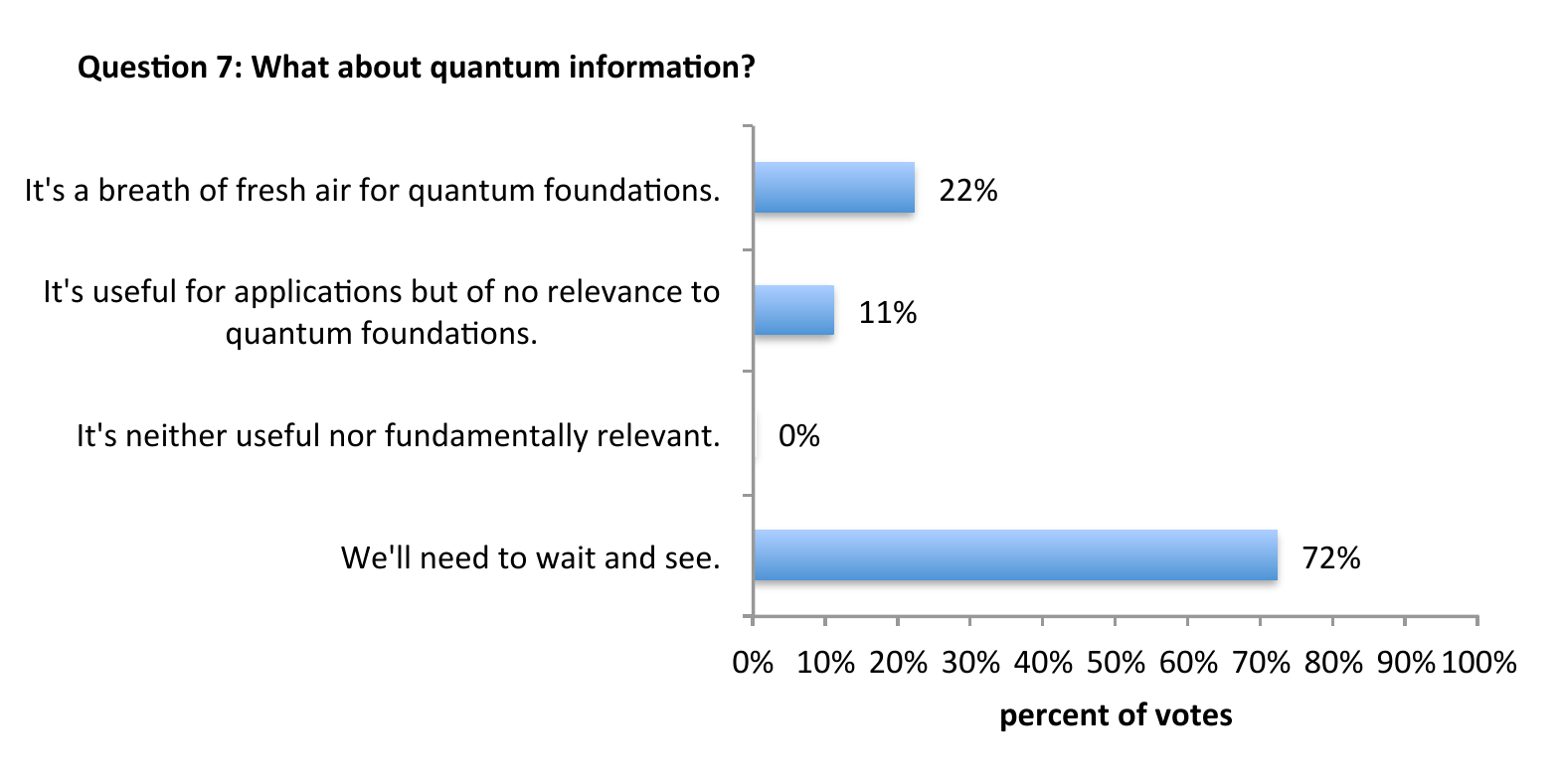}
\end{center}

In regard of quantum information, most participants again remained conservative and said that they have to wait and see. This is quite contrary to the study of Schlosshauer et al. \cite{Zeilinger}, where almost three fourth of the people called it a breath of fresh air for quantum foundations. Obviously, the answers to this question fiercely depend on the community being polled.

\begin{center}
\includegraphics[width=.7\linewidth]{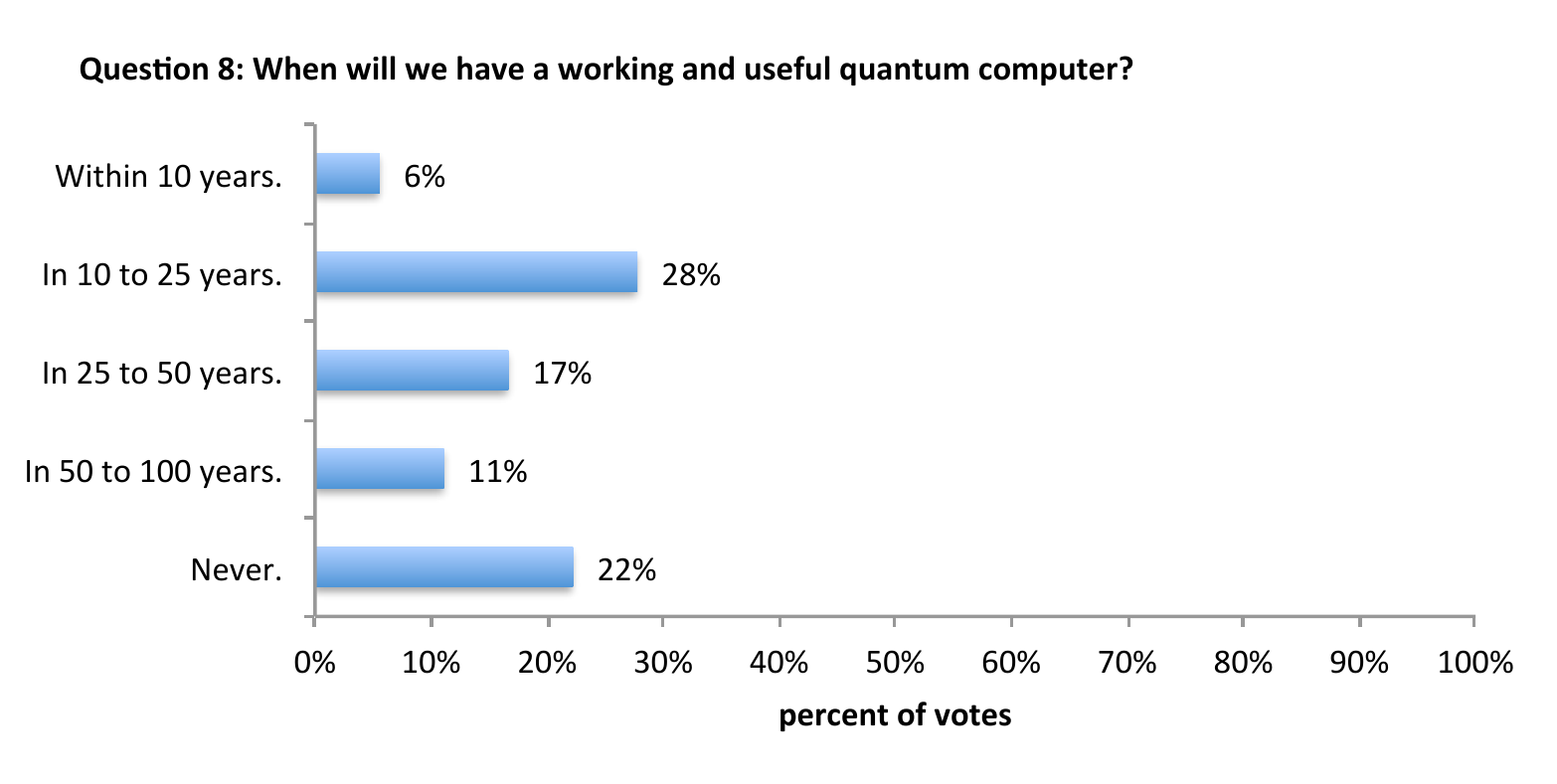}
\end{center}

There were many write-ins like "No idea." or "Define 'working' and 'useful'!". Given this indeterminacy, a good fifth does not believe that we will ever have such thing as a quantum computer and the rest of the answers peaks at "in 10 to 25 years" showing a general optimism regarding the progress in this field.

\begin{center}
\includegraphics[width=.7\linewidth]{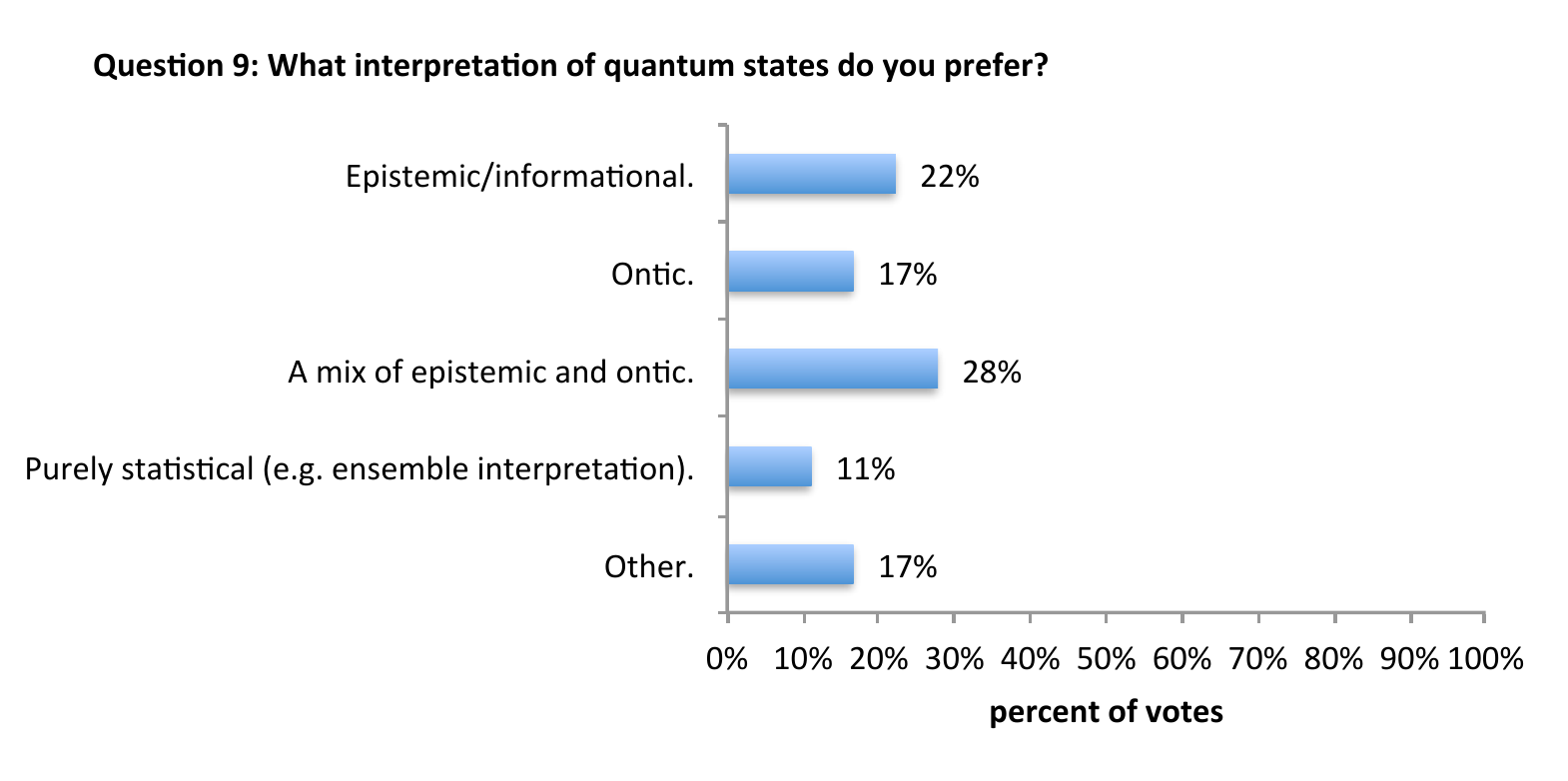}
\end{center}

All answers are scattered approximately uniformly between the possible answers. The purely statistical interpretation received considerably more votes than in the poll of Schlosshauer et al. \cite{Zeilinger}.

\begin{center}
\includegraphics[width=.7\linewidth]{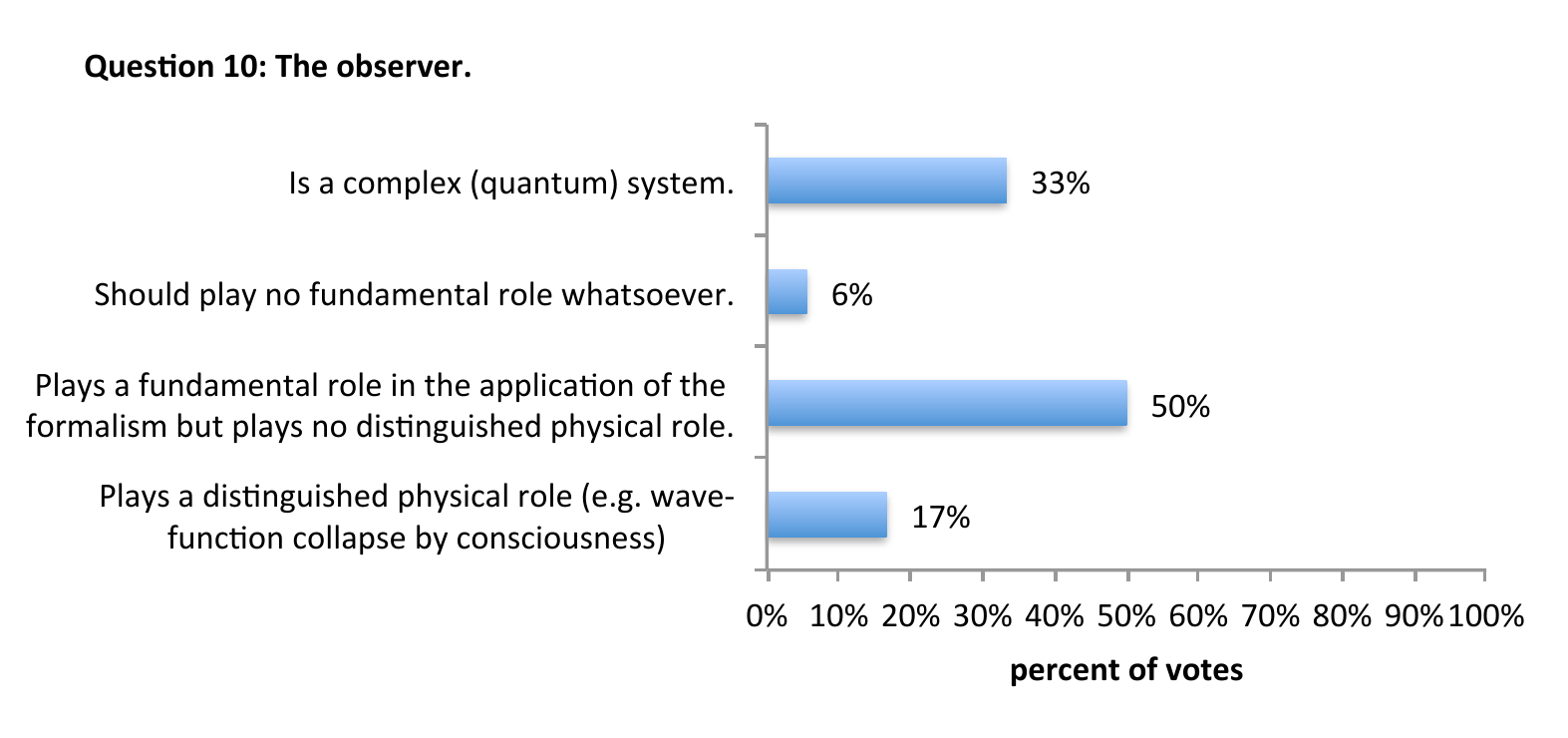}
\end{center}

The conference participants acknowledge some special role on behalf of the observer. One third see him as a complex (quantum) system and two thirds consider him either important for the application of the formalism or even fulfilling a distinguished physical role. Probably, the much discussed (quantum) mind-body problem, which was also hotly debated at the conference, played a part in this question.

\begin{center}
\includegraphics[width=.7\linewidth]{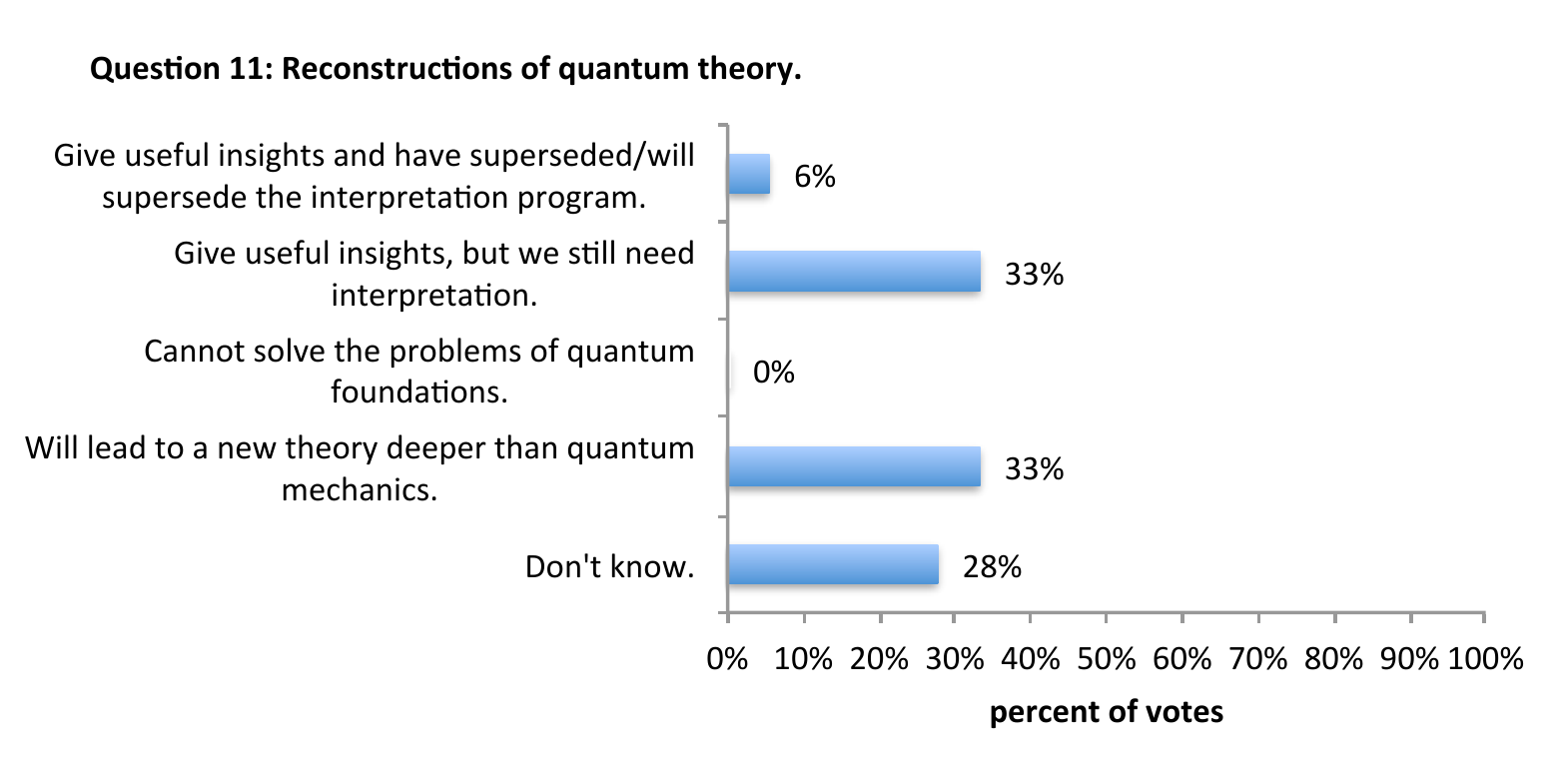}
\end{center}

A third of the respondents wait for a new, deeper theory than quantum mechanics and another third are convinced that reconstructions of quantum theory can, in principle, give useful insights, but do not relieve one from the need to find a proper interpretation.

\begin{center}
\includegraphics[width=.8\linewidth]{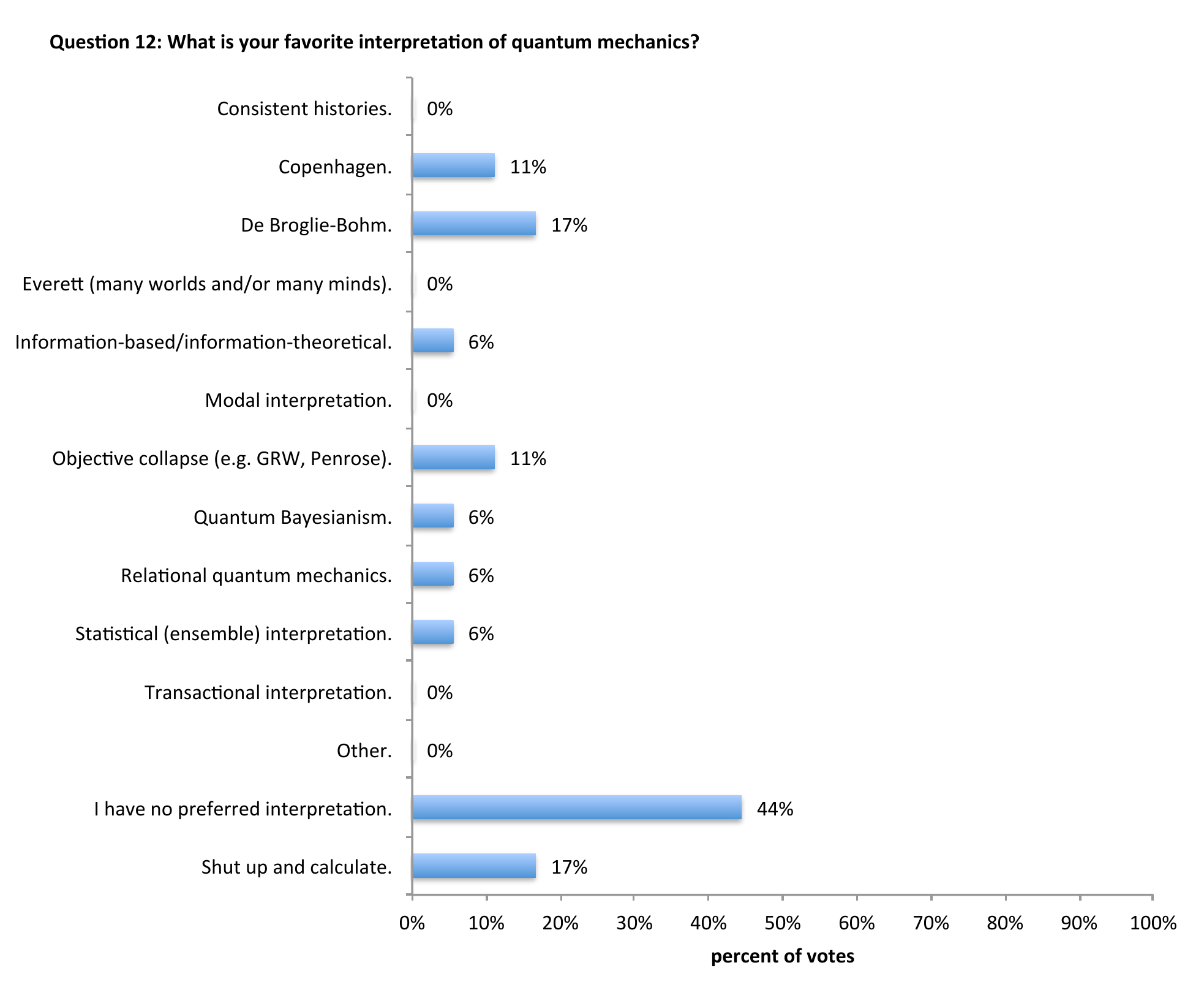}
\end{center}

This is a somehow surprising result. The elsewise sovereign Copenhagen interpretation loses ground against the de Broglie-Bohm interpretation \cite{deBroglie,Bohm1,Bohm2}. This is partly the influence of decided minorities in small populations, because the participants of the conference were all but representative of the whole physicists' community. Not surprisingly, the outcome is well different from the observed distribution by Tegmark \cite{Tegmark} or Schlosshauer et al. \cite{Zeilinger}.

Interestingly, the "Shut up and calculate." interpretation was the other big winner of this question.

\begin{center}
\includegraphics[width=.7\linewidth]{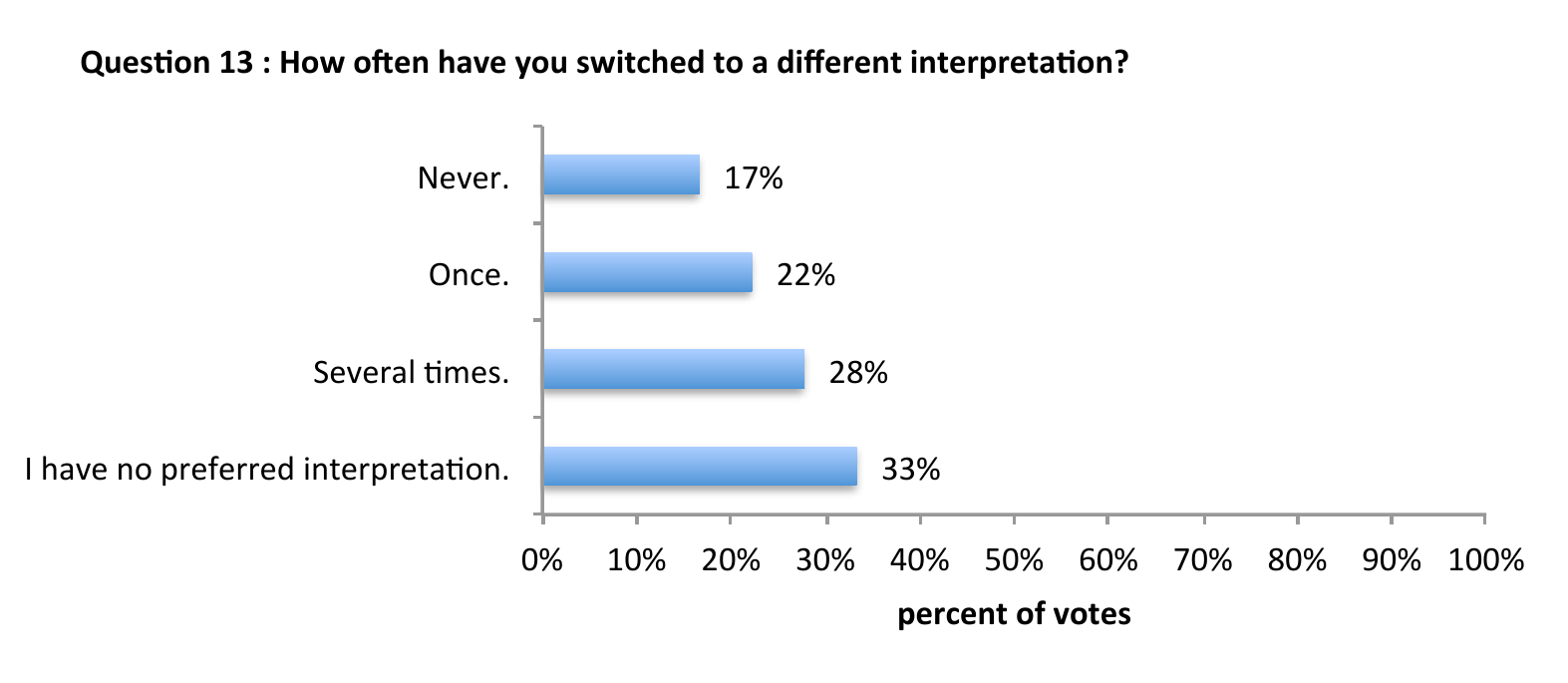}
\end{center}

The result shows an inclined distribution with a trend towards changing the favorite interpretation several times. A third of the participants do not even have a preferred interpretation. It is striking, however, that the answer "I have no preferred interpretation." did not receive the same number of votes as in Question 12, but at least there is a strong correlation (compare Sec. \ref{sec:corr} and especially Fig. \ref{fig:strongcor}). Also from the correlations of Fig. \ref{fig:strongcor}, one can deduce that the people who stated that they have switched interpretation several times were likely to state that the actual interpretation one assumes depends a lot on personal philosophical prejudice.

\begin{center}
\includegraphics[width=.7\linewidth]{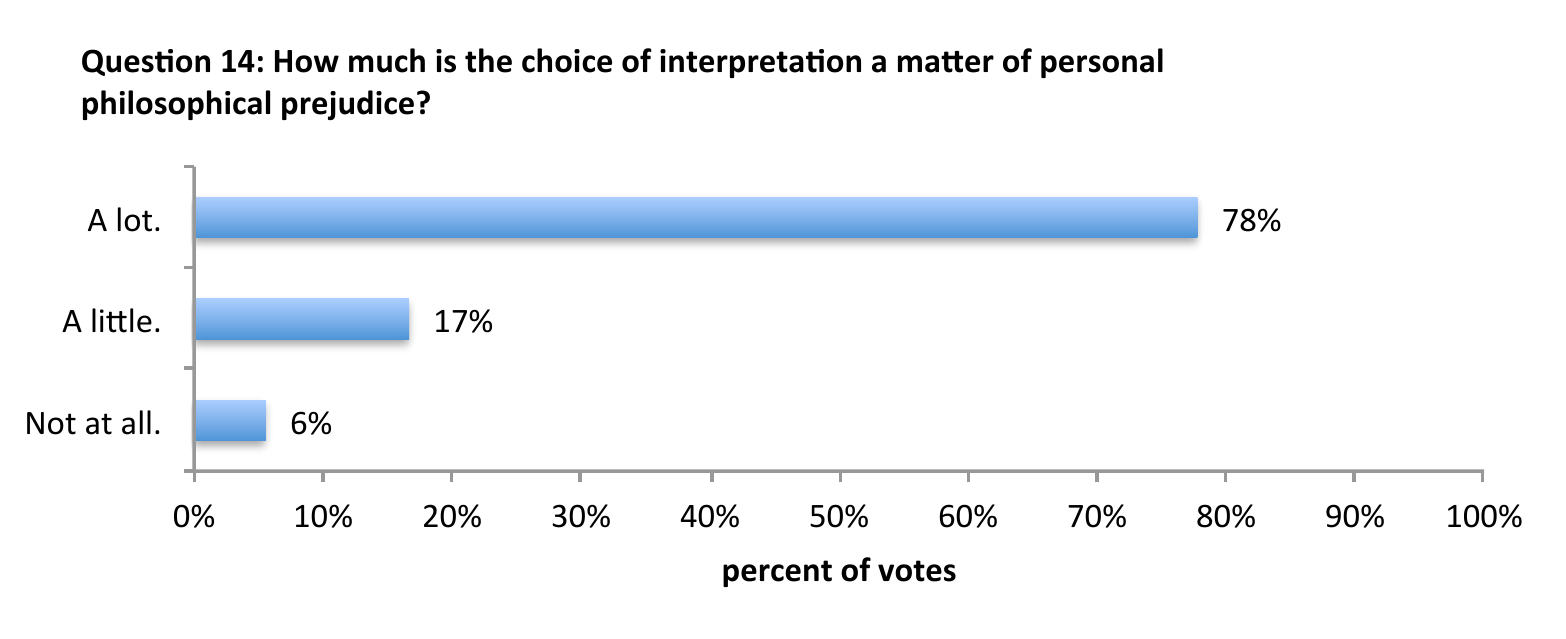}
\end{center}

Clearly, most of the participants think that the actual choice of interpretation is a matter of philosophical prejudice. Due to this high support, this answer could also be found quite often in the correlation table of Appendix \ref{sec:corrtab}. It seems that supporters of many different interpretations had to accept that in the absence of experimental evidence, the choice of interpretation is to a high degree a product of personal preference where reasons like philosophical beliefs or mathematical esthetics in the formulation of the theory play a decisive role.
The same trend was also observed by Schlosshauer et al. \cite{Zeilinger}.

\begin{center}
\includegraphics[width=.7\linewidth]{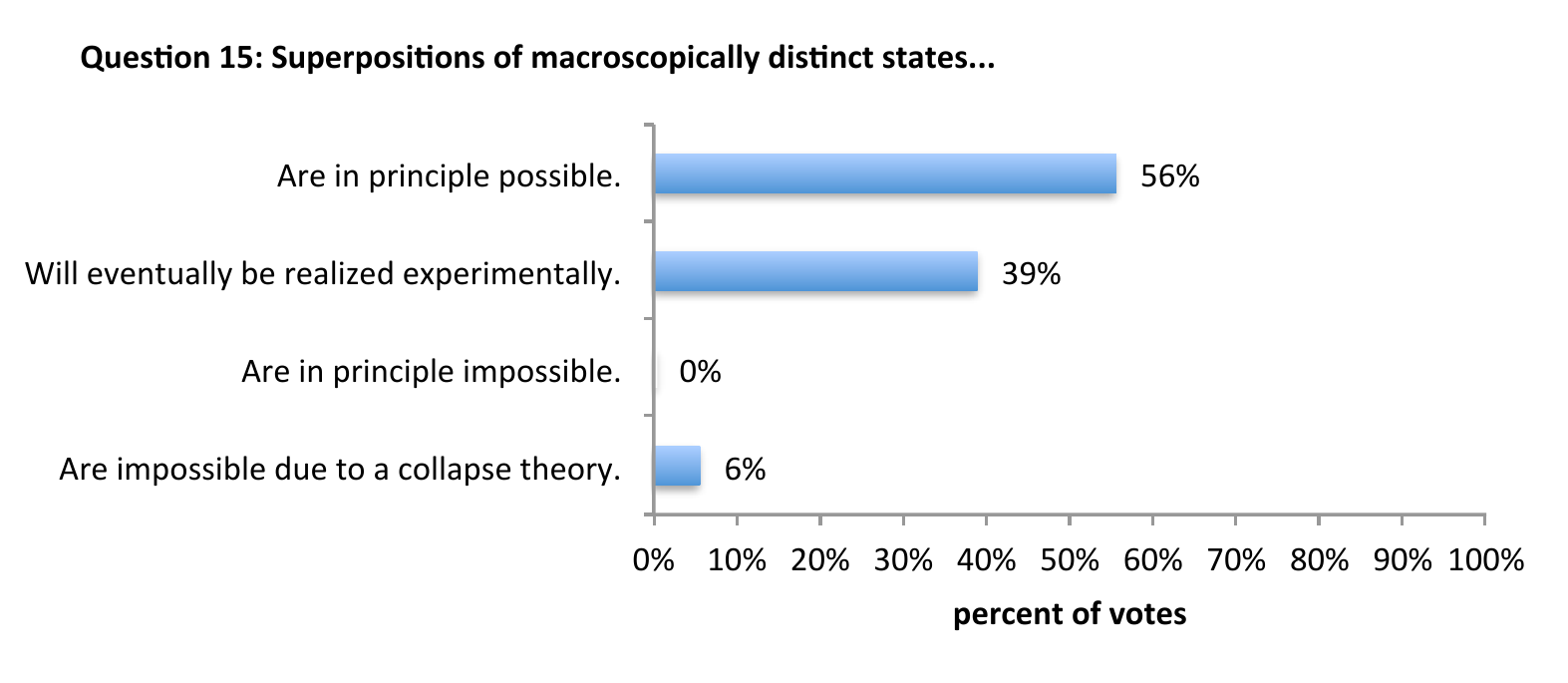}
\end{center}

More than half of the participants deemed the superposition of macroscopically distinct states in principal possible. A good third even thinks that they will be realized experimentally. This optimistic view is in congruency with the poll of Schlosshauer et al. \cite{Zeilinger}.

\begin{center}
\includegraphics[width=.7\linewidth]{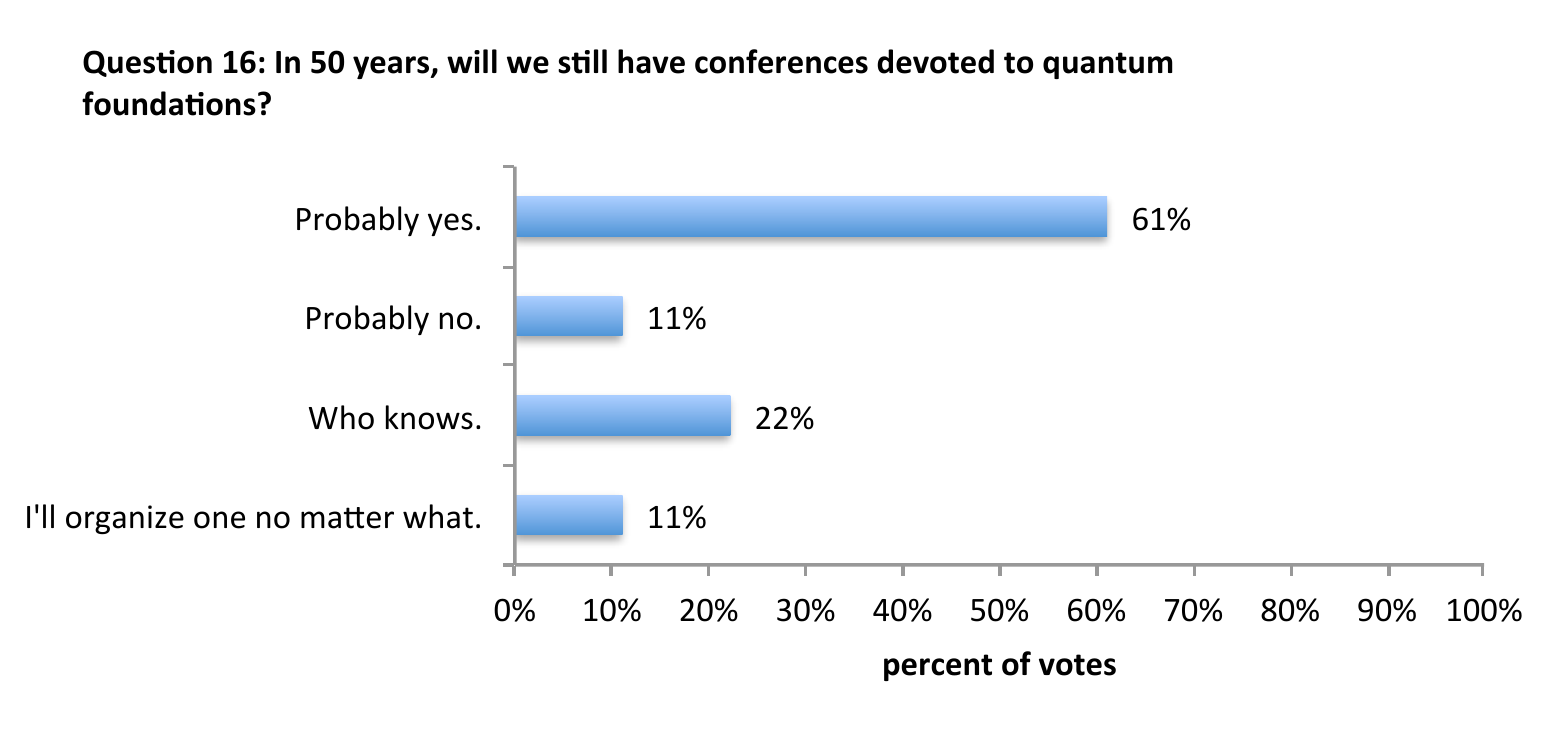}
\end{center}

At least the good spirits did not drop at the end of this conference, since there is a strong belief that there will be conferences even 50 years from now. Moreover, there seem to be already willing organizers (11\%)\dots

\section{\label{sec:corr}Correlations}

\begin{figure}[htp]
\begin{center}
\includegraphics[width=.9\linewidth]{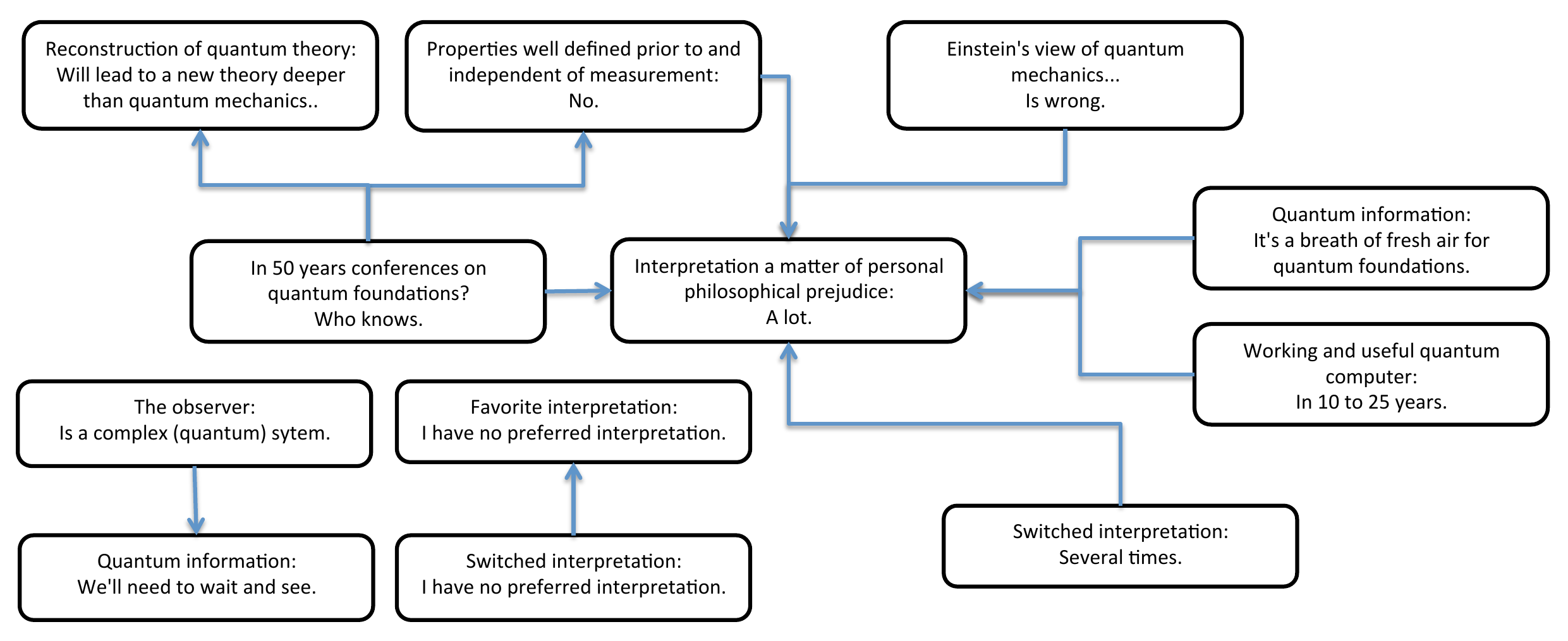}
\end{center}
\caption{Strong correlations between answers are shown. All answers have been checked by at least 4 participants. An arrow pointing from an answer A to answer B means that B is strongly correlated to A. For the definition of correlation, see Sec. \ref{sec:corr}.}
\label{fig:strongcor}
\end{figure}

It is quite enlightening to look for correlations in the answers. The criteria are chosen as with Schlosshauer et al. \cite{Zeilinger}. To suppress noise, any answer A to be correlated has to be checked by at least 4 participants. An answer B is called correlated if the fraction of votes that B received from A voters $\#(A\cap B)/\#A\geq T$ exceeds not only a threshold value $T$, but also the total fraction of votes for A by some gap value G, $\#(A\cap B)/\#A\geq (1+G) \cdot \#A/18$. The pair $(T,G)$ can be $(100\%,20\%)$ (strong correlation), $(75\%,15\%)$ (medium correlation) or $(50\%,10\%)$ (weak correlation).
Strong correlations alone are shown in Fig. \ref{fig:strongcor} and together with medium correlations in Fig. \ref{fig:medcor}. There are too many weak correlation to present them in a diagram, so they were added in Appendix \ref{sec:corrtab} for reference.
Note that due to the nature of the conditional probability, the above definition of correlation is not symmetric. As in the study of Schlosshauer et al. \cite{Zeilinger}, surprisingly few mutual correlations appear.

\begin{figure}[htp]
\begin{center}
\includegraphics[width=\linewidth]{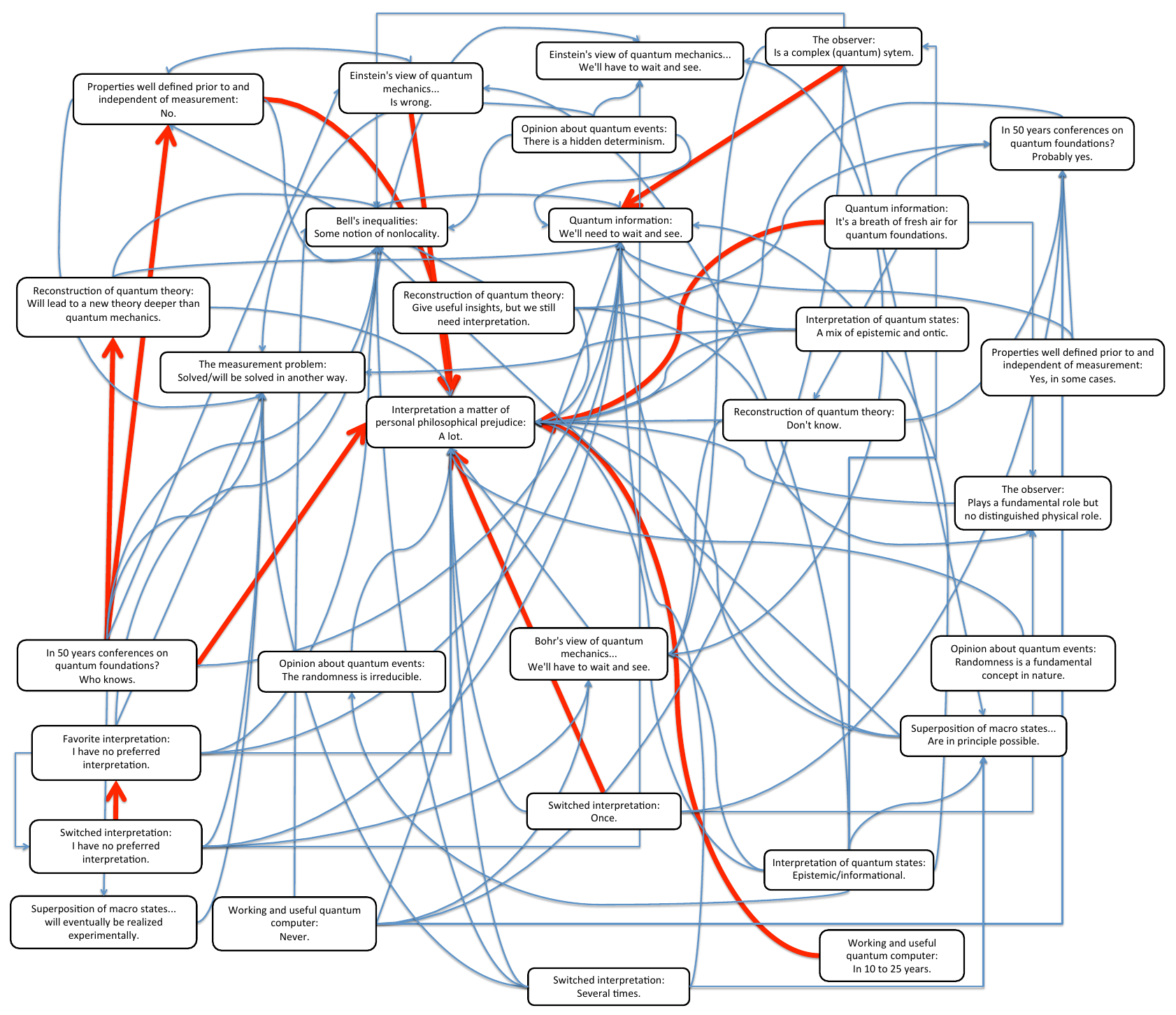}
\end{center}
\caption{Strong and medium correlations between answers are shown. All answers have been checked by at least 4 participants. A fat (red) arrow pointing from an answer A to answer B means that B is strongly correlated to A, a regular (blue) arrow means medium correlation. For the definition of correlation, see Sec. \ref{sec:corr}. Arrows always originate from centers of the sides of rectangles.}
\label{fig:medcor}
\end{figure}

\section{\label{sec:dis}Discussion}

In this section, a short discussion of the findings of this poll shall be given. There were some questions that found large support, most prominently the question whether the choice of interpretation is a matter of philosophical prejudice, where 78\% of the answers were strongly affirmative.
Also, there seems to be much confidence in experimental progress, because in question 15, a clear majority is convinced that the superposition of macroscopically distinct states is in principal possible (56\%) and 39\% think that it will be even realized experimentally.
The opinion that in 50 years, there will still be conferences devoted to quantum foundations shows that this problem is taken seriously by many people. Of course, there is always some biased error when a question like this is posed on a conference that deals explicitly with this problem.
Einstein's and Bohr's views found no unrestricted support as nobody could motivate himself to answer that their views are correct without limitation.

As became clear from comparison with the study of Schlosshauer et al. \cite{Zeilinger}, the views and opinions concerning the foundations of quantum mechanics depend on the basic population that is being polled, but nevertheless there were many parallels that could be drawn.

\section{\label{sec:sum}Summary}

In summary, another small snapshot of the views on the foundations of quantum theory was presented. While the details of such snapshots depend on the respondents who share their views, it was shown that there is still much controversy in many questions that lie at the very foundations of quantum mechanics. Although there is a general optimism towards experimental progress and theoretical effort, many people believe that there will still be conferences about this matter in 50 years. All the more, it is important to stay open to discussion given this multitude of convictions and interpretations.

\begin{acknowledgments}
The author wishes to thank J.~Kleiner for the invitation and the rest of the conference attendees for many interesting discussions and their disposition to participate in this poll on the foundations of quantum mechanics. Additionally, the kind support of the ``Studienstiftung des deutschen Volkes'' (German National Merit Foundation) is kindly acknowledged here.
\end{acknowledgments}

\appendix

\section{\label{sec:corrtab}Correlation table}

This table shows correlations between answers. Strong correlations are marked in red and labeled "3", medium correlations in orange with label "2" and weak correlations are yellow with label "1". In the first column, the total numbers of votes are listed. The definitions of strong, medium and weak correlations can be found in Sec. \ref{sec:corr}. In the first row, there is a short reminder of the corresponding question for reference. If "$A\Rightarrow B$", meaning B is correlated to A, then A questions can be found in rows and B questions in columns.\\[1em]

{\setlength{\tabcolsep}{0pt}
\begin{tabular}{cc*{30}{H{0.54cm}}}
\normalsize
 & & \mcn{3}{\tiny randomness} & \mcn{2}{\tiny props} & \mcn{2}{\tiny Einstein} & \mcn{1}{\tiny Bohr} & \mcn{1}{\tiny meas} & \mcn{1}{\tiny Bell} & \mcn{2}{\tiny Q info} & \mcn{2}{\tiny Q comp} & \mcn{2}{\tiny Q states} & \mcn{2}{\tiny observ} & \mcn{3}{\tiny reconstruct} & \mcn{1}{\tiny interp} & \mcn{3}{\tiny switch interp} & \mcn{1}{\tiny prej} & \mcn{2}{\tiny sup mac} & \mcn{2}{\tiny conference} \\
 \# & A\textbackslash B & 1b & 1c & 1d & 2b & 2c & 3b & 3e & 4e & 5c & 6c & 7a & 7d & 8b & 8e & 9a & 9c & 10a & 10c & 11b & 11d & 11e & 12m & 13b & 13c & 13d & 14a & 15a & 15b & 16a & 16c \\
5 & 1b & \textcolor{gelbbraun}{0} & \textcolor{gelbbraun}{0} & \textcolor{gelbbraun}{0} & \cellcolor{yellow}1 & \textcolor{gelbbraun}{0} & \textcolor{gelbbraun}{0} & \cellcolor{orange}2 & \textcolor{gelbbraun}{0} & \cellcolor{orange}2 & \cellcolor{orange}2 & \textcolor{gelbbraun}{0} & \cellcolor{orange}2 & \textcolor{gelbbraun}{0} & \textcolor{gelbbraun}{0} & \textcolor{gelbbraun}{0} & \textcolor{gelbbraun}{0} & \textcolor{gelbbraun}{0} & \textcolor{gelbbraun}{0} & \textcolor{gelbbraun}{0} & \textcolor{gelbbraun}{0} & \textcolor{gelbbraun}{0} & \cellcolor{yellow}1 & \textcolor{gelbbraun}{0} & \cellcolor{yellow}1 & \textcolor{gelbbraun}{0} & \cellcolor{yellow}1 & \textcolor{gelbbraun}{0} & \cellcolor{yellow}1 & \cellcolor{yellow}1 & \textcolor{gelbbraun}{0} \\
7 & 1c & \textcolor{gelbbraun}{0} & \textcolor{gelbbraun}{0} & \textcolor{gelbbraun}{0} & \textcolor{gelbbraun}{0} & \cellcolor{yellow}1 & \textcolor{gelbbraun}{0} & \cellcolor{yellow}1 & \cellcolor{yellow}1 & \cellcolor{yellow}1 & \cellcolor{yellow}1 & \textcolor{gelbbraun}{0} & \cellcolor{yellow}1 & \textcolor{gelbbraun}{0} & \textcolor{gelbbraun}{0} & \textcolor{gelbbraun}{0} & \textcolor{gelbbraun}{0} & \textcolor{gelbbraun}{0} & \cellcolor{yellow}1 & \cellcolor{yellow}1 & \cellcolor{yellow}1 & \textcolor{gelbbraun}{0} & \cellcolor{yellow}1 & \textcolor{gelbbraun}{0} & \textcolor{gelbbraun}{0} & \textcolor{gelbbraun}{0} & \cellcolor{orange}2 & \cellcolor{yellow}1 & \textcolor{gelbbraun}{0} & \cellcolor{yellow}1 & \textcolor{gelbbraun}{0} \\
7 & 1d & \textcolor{gelbbraun}{0} & \textcolor{gelbbraun}{0} & \textcolor{gelbbraun}{0} & \textcolor{gelbbraun}{0} & \cellcolor{yellow}1 & \cellcolor{yellow}1 & \textcolor{gelbbraun}{0} & \cellcolor{yellow}1 & \cellcolor{yellow}1 & \cellcolor{yellow}1 & \textcolor{gelbbraun}{0} & \cellcolor{yellow}1 & \textcolor{gelbbraun}{0} & \textcolor{gelbbraun}{0} & \textcolor{gelbbraun}{0} & \textcolor{gelbbraun}{0} & \textcolor{gelbbraun}{0} & \cellcolor{yellow}1 & \cellcolor{yellow}1 & \textcolor{gelbbraun}{0} & \textcolor{gelbbraun}{0} & \textcolor{gelbbraun}{0} & \textcolor{gelbbraun}{0} & \textcolor{gelbbraun}{0} & \textcolor{gelbbraun}{0} & \cellcolor{orange}2 & \cellcolor{yellow}1 & \textcolor{gelbbraun}{0} & \cellcolor{yellow}1 & \textcolor{gelbbraun}{0} \\
6 & 2b & \cellcolor{yellow}1 & \textcolor{gelbbraun}{0} & \textcolor{gelbbraun}{0} & \textcolor{gelbbraun}{0} & \textcolor{gelbbraun}{0} & \textcolor{gelbbraun}{0} & \cellcolor{yellow}1 & \cellcolor{yellow}1 & \cellcolor{yellow}1 & \cellcolor{yellow}1 & \textcolor{gelbbraun}{0} & \cellcolor{orange}2 & \textcolor{gelbbraun}{0} & \textcolor{gelbbraun}{0} & \textcolor{gelbbraun}{0} & \cellcolor{yellow}1 & \textcolor{gelbbraun}{0} & \cellcolor{yellow}1 & \textcolor{gelbbraun}{0} & \textcolor{gelbbraun}{0} & \textcolor{gelbbraun}{0} & \cellcolor{yellow}1 & \textcolor{gelbbraun}{0} & \textcolor{gelbbraun}{0} & \textcolor{gelbbraun}{0} & \cellcolor{yellow}1 & \cellcolor{yellow}1 & \cellcolor{yellow}1 & \cellcolor{orange}2 & \textcolor{gelbbraun}{0} \\
8 & 2c & \textcolor{gelbbraun}{0} & \cellcolor{yellow}1 & \cellcolor{yellow}1 & \textcolor{gelbbraun}{0} & \textcolor{gelbbraun}{0} & \cellcolor{orange}2 & \textcolor{gelbbraun}{0} & \cellcolor{yellow}1 & \cellcolor{orange}2 & \cellcolor{orange}2 & \textcolor{gelbbraun}{0} & \cellcolor{yellow}1 & \textcolor{gelbbraun}{0} & \textcolor{gelbbraun}{0} & \textcolor{gelbbraun}{0} & \textcolor{gelbbraun}{0} & \textcolor{gelbbraun}{0} & \cellcolor{yellow}1 & \cellcolor{yellow}1 & \cellcolor{yellow}1 & \textcolor{gelbbraun}{0} & \textcolor{gelbbraun}{0} & \textcolor{gelbbraun}{0} & \textcolor{gelbbraun}{0} & \textcolor{gelbbraun}{0} & \cellcolor{red}3 & \cellcolor{yellow}1 & \textcolor{gelbbraun}{0} & \cellcolor{yellow}1 & \cellcolor{yellow}1 \\
7 & 3b & \textcolor{gelbbraun}{0} & \textcolor{gelbbraun}{0} & \cellcolor{yellow}1 & \textcolor{gelbbraun}{0} & \cellcolor{orange}2 & \textcolor{gelbbraun}{0} & \textcolor{gelbbraun}{0} & \cellcolor{yellow}1 & \cellcolor{yellow}1 & \cellcolor{yellow}1 & \textcolor{gelbbraun}{0} & \cellcolor{yellow}1 & \textcolor{gelbbraun}{0} & \textcolor{gelbbraun}{0} & \textcolor{gelbbraun}{0} & \textcolor{gelbbraun}{0} & \textcolor{gelbbraun}{0} & \cellcolor{yellow}1 & \cellcolor{yellow}1 & \textcolor{gelbbraun}{0} & \textcolor{gelbbraun}{0} & \textcolor{gelbbraun}{0} & \textcolor{gelbbraun}{0} & \textcolor{gelbbraun}{0} & \textcolor{gelbbraun}{0} & \cellcolor{red}3 & \cellcolor{yellow}1 & \textcolor{gelbbraun}{0} & \cellcolor{yellow}1 & \textcolor{gelbbraun}{0} \\
10 & 3e & \textcolor{gelbbraun}{0} & \textcolor{gelbbraun}{0} & \textcolor{gelbbraun}{0} & \textcolor{gelbbraun}{0} & \textcolor{gelbbraun}{0} & \textcolor{gelbbraun}{0} & \textcolor{gelbbraun}{0} & \cellcolor{yellow}1 & \textcolor{gelbbraun}{0} & \textcolor{gelbbraun}{0} & \textcolor{gelbbraun}{0} & \cellcolor{yellow}1 & \textcolor{gelbbraun}{0} & \textcolor{gelbbraun}{0} & \textcolor{gelbbraun}{0} & \textcolor{gelbbraun}{0} & \textcolor{gelbbraun}{0} & \textcolor{gelbbraun}{0} & \textcolor{gelbbraun}{0} & \textcolor{gelbbraun}{0} & \textcolor{gelbbraun}{0} & \cellcolor{yellow}1 & \textcolor{gelbbraun}{0} & \textcolor{gelbbraun}{0} & \textcolor{gelbbraun}{0} & \cellcolor{yellow}1 & \textcolor{gelbbraun}{0} & \textcolor{gelbbraun}{0} & \textcolor{gelbbraun}{0} & \textcolor{gelbbraun}{0} \\
11 & 4e & \textcolor{gelbbraun}{0} & \textcolor{gelbbraun}{0} & \textcolor{gelbbraun}{0} & \textcolor{gelbbraun}{0} & \textcolor{gelbbraun}{0} & \textcolor{gelbbraun}{0} & \textcolor{gelbbraun}{0} & \textcolor{gelbbraun}{0} & \textcolor{gelbbraun}{0} & \textcolor{gelbbraun}{0} & \textcolor{gelbbraun}{0} & \cellcolor{yellow}1 & \textcolor{gelbbraun}{0} & \textcolor{gelbbraun}{0} & \textcolor{gelbbraun}{0} & \textcolor{gelbbraun}{0} & \textcolor{gelbbraun}{0} & \textcolor{gelbbraun}{0} & \textcolor{gelbbraun}{0} & \textcolor{gelbbraun}{0} & \textcolor{gelbbraun}{0} & \textcolor{gelbbraun}{0} & \textcolor{gelbbraun}{0} & \textcolor{gelbbraun}{0} & \textcolor{gelbbraun}{0} & \cellcolor{orange}2 & \textcolor{gelbbraun}{0} & \textcolor{gelbbraun}{0} & \cellcolor{yellow}1 & \textcolor{gelbbraun}{0} \\
11 & 5c & \textcolor{gelbbraun}{0} & \textcolor{gelbbraun}{0} & \textcolor{gelbbraun}{0} & \textcolor{gelbbraun}{0} & \textcolor{gelbbraun}{0} & \textcolor{gelbbraun}{0} & \textcolor{gelbbraun}{0} & \textcolor{gelbbraun}{0} & \textcolor{gelbbraun}{0} & \cellcolor{yellow}1 & \textcolor{gelbbraun}{0} & \cellcolor{yellow}1 & \textcolor{gelbbraun}{0} & \textcolor{gelbbraun}{0} & \textcolor{gelbbraun}{0} & \textcolor{gelbbraun}{0} & \textcolor{gelbbraun}{0} & \textcolor{gelbbraun}{0} & \textcolor{gelbbraun}{0} & \textcolor{gelbbraun}{0} & \textcolor{gelbbraun}{0} & \textcolor{gelbbraun}{0} & \textcolor{gelbbraun}{0} & \textcolor{gelbbraun}{0} & \textcolor{gelbbraun}{0} & \cellcolor{yellow}1 & \textcolor{gelbbraun}{0} & \textcolor{gelbbraun}{0} & \textcolor{gelbbraun}{0} & \textcolor{gelbbraun}{0} \\
12 & 6c & \textcolor{gelbbraun}{0} & \textcolor{gelbbraun}{0} & \textcolor{gelbbraun}{0} & \textcolor{gelbbraun}{0} & \textcolor{gelbbraun}{0} & \textcolor{gelbbraun}{0} & \textcolor{gelbbraun}{0} & \textcolor{gelbbraun}{0} & \textcolor{gelbbraun}{0} & \textcolor{gelbbraun}{0} & \textcolor{gelbbraun}{0} & \cellcolor{orange}2 & \textcolor{gelbbraun}{0} & \textcolor{gelbbraun}{0} & \textcolor{gelbbraun}{0} & \textcolor{gelbbraun}{0} & \textcolor{gelbbraun}{0} & \textcolor{gelbbraun}{0} & \textcolor{gelbbraun}{0} & \textcolor{gelbbraun}{0} & \textcolor{gelbbraun}{0} & \textcolor{gelbbraun}{0} & \textcolor{gelbbraun}{0} & \textcolor{gelbbraun}{0} & \textcolor{gelbbraun}{0} & \cellcolor{yellow}1 & \textcolor{gelbbraun}{0} & \textcolor{gelbbraun}{0} & \textcolor{gelbbraun}{0} & \textcolor{gelbbraun}{0} \\
4 & 7a & \textcolor{gelbbraun}{0} & \textcolor{gelbbraun}{0} & \cellcolor{yellow}1 & \textcolor{gelbbraun}{0} & \cellcolor{yellow}1 & \cellcolor{yellow}1 & \cellcolor{yellow}1 & \cellcolor{yellow}1 & \cellcolor{yellow}1 & \textcolor{gelbbraun}{0} & \textcolor{gelbbraun}{0} & \textcolor{gelbbraun}{0} & \textcolor{gelbbraun}{0} & \textcolor{gelbbraun}{0} & \textcolor{gelbbraun}{0} & \textcolor{gelbbraun}{0} & \textcolor{gelbbraun}{0} & \cellcolor{orange}2 & \cellcolor{yellow}1 & \textcolor{gelbbraun}{0} & \cellcolor{orange}2 & \textcolor{gelbbraun}{0} & \cellcolor{yellow}1 & \textcolor{gelbbraun}{0} & \textcolor{gelbbraun}{0} & \cellcolor{red}3 & \textcolor{gelbbraun}{0} & \cellcolor{yellow}1 & \cellcolor{orange}2 & \textcolor{gelbbraun}{0} \\
13 & 7d & \textcolor{gelbbraun}{0} & \textcolor{gelbbraun}{0} & \textcolor{gelbbraun}{0} & \textcolor{gelbbraun}{0} & \textcolor{gelbbraun}{0} & \textcolor{gelbbraun}{0} & \textcolor{gelbbraun}{0} & \textcolor{gelbbraun}{0} & \textcolor{gelbbraun}{0} & \cellcolor{orange}2 & \textcolor{gelbbraun}{0} & \textcolor{gelbbraun}{0} & \textcolor{gelbbraun}{0} & \textcolor{gelbbraun}{0} & \textcolor{gelbbraun}{0} & \textcolor{gelbbraun}{0} & \textcolor{gelbbraun}{0} & \textcolor{gelbbraun}{0} & \textcolor{gelbbraun}{0} & \textcolor{gelbbraun}{0} & \textcolor{gelbbraun}{0} & \textcolor{gelbbraun}{0} & \textcolor{gelbbraun}{0} & \textcolor{gelbbraun}{0} & \textcolor{gelbbraun}{0} & \textcolor{gelbbraun}{0} & \textcolor{gelbbraun}{0} & \textcolor{gelbbraun}{0} & \textcolor{gelbbraun}{0} & \textcolor{gelbbraun}{0} \\
5& 8b & \textcolor{gelbbraun}{0} & \cellcolor{yellow}1 & \cellcolor{yellow}1 & \textcolor{gelbbraun}{0} & \cellcolor{yellow}1 & \cellcolor{yellow}1 & \textcolor{gelbbraun}{0} & \cellcolor{yellow}1 & \textcolor{gelbbraun}{0} & \textcolor{gelbbraun}{0} & \textcolor{gelbbraun}{0} & \cellcolor{yellow}1 & \textcolor{gelbbraun}{0} & \textcolor{gelbbraun}{0} & \textcolor{gelbbraun}{0} & \textcolor{gelbbraun}{0} & \textcolor{gelbbraun}{0} & \cellcolor{yellow}1 & \cellcolor{yellow}1 & \textcolor{gelbbraun}{0} & \textcolor{gelbbraun}{0} & \cellcolor{yellow}1 & \textcolor{gelbbraun}{0} & \textcolor{gelbbraun}{0} & \textcolor{gelbbraun}{0} & \cellcolor{red}3 & \cellcolor{yellow}1 & \textcolor{gelbbraun}{0} & \cellcolor{yellow}1 & \textcolor{gelbbraun}{0} \\
4 & 8e & \cellcolor{yellow}1 & \cellcolor{yellow}1 & \textcolor{gelbbraun}{0} & \textcolor{gelbbraun}{0} & \textcolor{gelbbraun}{0} & \textcolor{gelbbraun}{0} & \cellcolor{yellow}1 & \cellcolor{orange}2 & \cellcolor{yellow}1 & \cellcolor{orange}2 & \textcolor{gelbbraun}{0} & \cellcolor{orange}2 & \textcolor{gelbbraun}{0} & \textcolor{gelbbraun}{0} & \cellcolor{yellow}1 & \textcolor{gelbbraun}{0} & \cellcolor{orange}2 & \textcolor{gelbbraun}{0} & \textcolor{gelbbraun}{0} & \textcolor{gelbbraun}{0} & \textcolor{gelbbraun}{0} & \textcolor{gelbbraun}{0} & \cellcolor{yellow}1 & \textcolor{gelbbraun}{0} & \textcolor{gelbbraun}{0} & \cellcolor{yellow}1 & \cellcolor{yellow}1 & \cellcolor{yellow}1 & \cellcolor{orange}2 & \textcolor{gelbbraun}{0} \\
4 & 9a & \textcolor{gelbbraun}{0} & \cellcolor{orange}2 & \cellcolor{yellow}1 & \textcolor{gelbbraun}{0} & \cellcolor{yellow}1 & \cellcolor{orange}2 & \textcolor{gelbbraun}{0} & \cellcolor{orange}2 & \textcolor{gelbbraun}{0} & \cellcolor{yellow}1 & \textcolor{gelbbraun}{0} & \cellcolor{orange}2 & \cellcolor{yellow}1 & \cellcolor{yellow}1 & \textcolor{gelbbraun}{0} & \textcolor{gelbbraun}{0} & \cellcolor{orange}2 & \cellcolor{yellow}1 & \cellcolor{yellow}1 & \cellcolor{yellow}1 & \cellcolor{yellow}1 & \textcolor{gelbbraun}{0} & \cellcolor{yellow}1 & \textcolor{gelbbraun}{0} & \textcolor{gelbbraun}{0} & \cellcolor{yellow}1 & \cellcolor{orange}2 & \textcolor{gelbbraun}{0} & \cellcolor{yellow}1 & \textcolor{gelbbraun}{0} \\
5 & 9c & \textcolor{gelbbraun}{0} & \textcolor{gelbbraun}{0} & \textcolor{gelbbraun}{0} & \cellcolor{yellow}1 & \textcolor{gelbbraun}{0} & \textcolor{gelbbraun}{0} & \cellcolor{orange}2 & \cellcolor{orange}2 & \cellcolor{orange}2 & \cellcolor{yellow}1 & \textcolor{gelbbraun}{0} & \cellcolor{orange}2 & \textcolor{gelbbraun}{0} & \textcolor{gelbbraun}{0} & \textcolor{gelbbraun}{0} & \textcolor{gelbbraun}{0} & \textcolor{gelbbraun}{0} & \cellcolor{yellow}1 & \textcolor{gelbbraun}{0} & \cellcolor{yellow}1 & \textcolor{gelbbraun}{0} & \cellcolor{yellow}1 & \textcolor{gelbbraun}{0} & \textcolor{gelbbraun}{0} & \textcolor{gelbbraun}{0} & \cellcolor{orange}2 & \cellcolor{yellow}1 & \textcolor{gelbbraun}{0} & \cellcolor{yellow}1 & \textcolor{gelbbraun}{0} \\
6 & 10a & \textcolor{gelbbraun}{0} & \textcolor{gelbbraun}{0} & \cellcolor{yellow}1 & \textcolor{gelbbraun}{0} & \textcolor{gelbbraun}{0} & \cellcolor{yellow}1 & \textcolor{gelbbraun}{0} & \cellcolor{orange}2 & \textcolor{gelbbraun}{0} & \cellcolor{orange}2 & \textcolor{gelbbraun}{0} & \cellcolor{red}3 & \textcolor{gelbbraun}{0} & \cellcolor{yellow}1 & \cellcolor{yellow}1 & \textcolor{gelbbraun}{0} & \textcolor{gelbbraun}{0} & \textcolor{gelbbraun}{0} & \textcolor{gelbbraun}{0} & \textcolor{gelbbraun}{0} & \textcolor{gelbbraun}{0} & \textcolor{gelbbraun}{0} & \textcolor{gelbbraun}{0} & \textcolor{gelbbraun}{0} & \textcolor{gelbbraun}{0} & \cellcolor{orange}2 & \cellcolor{yellow}1 & \textcolor{gelbbraun}{0} & \cellcolor{yellow}1 & \textcolor{gelbbraun}{0} \\
9 & 10c & \textcolor{gelbbraun}{0} & \textcolor{gelbbraun}{0} & \textcolor{gelbbraun}{0} & \textcolor{gelbbraun}{0} & \cellcolor{yellow}1 & \cellcolor{yellow}1 & \textcolor{gelbbraun}{0} & \textcolor{gelbbraun}{0} & \cellcolor{yellow}1 & \cellcolor{yellow}1 & \textcolor{gelbbraun}{0} & \cellcolor{yellow}1 & \textcolor{gelbbraun}{0} & \textcolor{gelbbraun}{0} & \textcolor{gelbbraun}{0} & \textcolor{gelbbraun}{0} & \textcolor{gelbbraun}{0} & \textcolor{gelbbraun}{0} & \cellcolor{yellow}1 & \textcolor{gelbbraun}{0} & \textcolor{gelbbraun}{0} & \textcolor{gelbbraun}{0} & \textcolor{gelbbraun}{0} & \textcolor{gelbbraun}{0} & \textcolor{gelbbraun}{0} & \cellcolor{yellow}1 & \cellcolor{yellow}1 & \textcolor{gelbbraun}{0} & \cellcolor{yellow}1 & \textcolor{gelbbraun}{0} \\
6 & 11b & \textcolor{gelbbraun}{0} & \cellcolor{yellow}1 & \cellcolor{yellow}1 & \textcolor{gelbbraun}{0} &\cellcolor{orange}2 & \cellcolor{yellow}1 & \textcolor{gelbbraun}{0} & \cellcolor{yellow}1 & \cellcolor{yellow}1 & \cellcolor{yellow}1 & \textcolor{gelbbraun}{0} & \cellcolor{yellow}1 & \cellcolor{yellow}1 & \textcolor{gelbbraun}{0} & \textcolor{gelbbraun}{0} & \textcolor{gelbbraun}{0} & \textcolor{gelbbraun}{0} & \cellcolor{orange}2 & \textcolor{gelbbraun}{0} & \textcolor{gelbbraun}{0} & \textcolor{gelbbraun}{0} & \cellcolor{yellow}1 & \textcolor{gelbbraun}{0} & \textcolor{gelbbraun}{0} & \cellcolor{yellow}1 & \cellcolor{yellow}1 & \textcolor{gelbbraun}{0} & \textcolor{gelbbraun}{0} & \cellcolor{orange}2 & \textcolor{gelbbraun}{0} \\
6 & 11d & \textcolor{gelbbraun}{0} & \cellcolor{yellow}1 & \textcolor{gelbbraun}{0} & \textcolor{gelbbraun}{0} & \cellcolor{yellow}1 & \cellcolor{yellow}1 & \cellcolor{yellow}1 & \cellcolor{yellow}1 & \cellcolor{yellow}1 & \cellcolor{orange}2 & \textcolor{gelbbraun}{0} &\cellcolor{orange}2 &  \textcolor{gelbbraun}{0} & \textcolor{gelbbraun}{0} & \textcolor{gelbbraun}{0} & \cellcolor{yellow}1 & \textcolor{gelbbraun}{0} & \cellcolor{yellow}1 & \textcolor{gelbbraun}{0} & \textcolor{gelbbraun}{0} & \textcolor{gelbbraun}{0} & \textcolor{gelbbraun}{0} & \textcolor{gelbbraun}{0} & \cellcolor{yellow}1 & \textcolor{gelbbraun}{0} & \cellcolor{orange}2 & \cellcolor{yellow}1 & \cellcolor{yellow}1 & \textcolor{gelbbraun}{0} & \cellcolor{yellow}1 \\
5 & 11e & \textcolor{gelbbraun}{0} & \textcolor{gelbbraun}{0} & \cellcolor{yellow}1 & \textcolor{gelbbraun}{0} & \textcolor{gelbbraun}{0} & \textcolor{gelbbraun}{0} & \cellcolor{yellow}1 & \cellcolor{orange}2 & \textcolor{gelbbraun}{0} & \textcolor{gelbbraun}{0} & \textcolor{gelbbraun}{0} & \cellcolor{yellow}1 & \textcolor{gelbbraun}{0} & \textcolor{gelbbraun}{0} & \textcolor{gelbbraun}{0} & \textcolor{gelbbraun}{0} & \textcolor{gelbbraun}{0} & \textcolor{gelbbraun}0 & \textcolor{gelbbraun}{0} & \textcolor{gelbbraun}{0} & \textcolor{gelbbraun}{0} & \textcolor{gelbbraun}{0} & \textcolor{gelbbraun}{0} & \textcolor{gelbbraun}{0} & \textcolor{gelbbraun}0 & \cellcolor{orange}2 & \cellcolor{yellow}1 & \textcolor{gelbbraun}{0} & \cellcolor{orange}2 & \textcolor{gelbbraun}{0} \\
8 & 12m & \textcolor{gelbbraun}{0} & \cellcolor{yellow}1 & \textcolor{gelbbraun}{0} & \textcolor{gelbbraun}{0} & \textcolor{gelbbraun}{0} & \textcolor{gelbbraun}{0} & \cellcolor{orange}2 & \cellcolor{yellow}1 & \cellcolor{orange}2 & \cellcolor{orange}2 & \textcolor{gelbbraun}{0} & \cellcolor{orange}2 & \textcolor{gelbbraun}{0} & \textcolor{gelbbraun}{0} & \textcolor{gelbbraun}{0} & \textcolor{gelbbraun}{0} & \textcolor{gelbbraun}{0} & \cellcolor{yellow}1 & \textcolor{gelbbraun}{0} & \textcolor{gelbbraun}{0} & \textcolor{gelbbraun}{0} & \textcolor{gelbbraun}{0} & \textcolor{gelbbraun}{0} & \textcolor{gelbbraun}{0} & \cellcolor{orange}2 & \cellcolor{orange}2 & \cellcolor{yellow}1 & \textcolor{gelbbraun}{0} & \cellcolor{yellow}1 & \textcolor{gelbbraun}{0} \\
4 & 13b & \textcolor{gelbbraun}{0} & \cellcolor{yellow}1 & \textcolor{gelbbraun}{0} & \cellcolor{yellow}1 & \cellcolor{yellow}1 & \cellcolor{yellow}1 & \textcolor{gelbbraun}{0} & \cellcolor{yellow}1 & \textcolor{gelbbraun}{0} & \cellcolor{yellow}1 & \cellcolor{yellow}1 & \cellcolor{yellow}1 & \textcolor{gelbbraun}{0} & \cellcolor{yellow}1 & \cellcolor{yellow}1 & \textcolor{gelbbraun}{0} & \cellcolor{yellow}1 & \cellcolor{orange}2 & \cellcolor{yellow}1 & \textcolor{gelbbraun}{0} & \cellcolor{yellow}1 & \textcolor{gelbbraun}{0} & \textcolor{gelbbraun}{0} & \textcolor{gelbbraun}{0} & \textcolor{gelbbraun}{0} & \cellcolor{orange}2 & \cellcolor{yellow}1 & \cellcolor{yellow}1 & \cellcolor{orange}2 & \textcolor{gelbbraun}{0} \\
5 & 13c & \cellcolor{yellow}1 & \textcolor{gelbbraun}{0} & \textcolor{gelbbraun}{0} & \textcolor{gelbbraun}{0} & \textcolor{gelbbraun}{0} & \textcolor{gelbbraun}{0} & \cellcolor{yellow}1 & \textcolor{gelbbraun}{0} & \cellcolor{orange}2 & \cellcolor{orange}2 & \textcolor{gelbbraun}{0} & \cellcolor{orange}2 & \textcolor{gelbbraun}{0} & \textcolor{gelbbraun}{0} & \textcolor{gelbbraun}{0} & \textcolor{gelbbraun}{0} & \textcolor{gelbbraun}{0} & \cellcolor{yellow}1 & \textcolor{gelbbraun}{0} & \cellcolor{yellow}1 & \textcolor{gelbbraun}{0} & \textcolor{gelbbraun}{0} & \textcolor{gelbbraun}{0} & \textcolor{gelbbraun}{0} & \textcolor{gelbbraun}{0} & \cellcolor{orange}2 & \cellcolor{orange}2 & \textcolor{gelbbraun}{0} & \textcolor{gelbbraun}{0} & \textcolor{gelbbraun}{0} \\
6 & 13d & \textcolor{gelbbraun}{0} & \cellcolor{yellow}1 & \cellcolor{yellow}1 & \textcolor{gelbbraun}{0} & \cellcolor{yellow}1 & \textcolor{gelbbraun}{0} & \cellcolor{orange}2 & \cellcolor{orange}2 & \cellcolor{orange}2 & \cellcolor{yellow}1 & \textcolor{gelbbraun}{0} & \cellcolor{orange}2 &  \textcolor{gelbbraun}{0} & \textcolor{gelbbraun}{0} & \textcolor{gelbbraun}{0} & \textcolor{gelbbraun}{0} & \textcolor{gelbbraun}{0} & \textcolor{gelbbraun}{0} & \cellcolor{yellow}1 & \textcolor{gelbbraun}{0} & \textcolor{gelbbraun}{0} & \cellcolor{red}3 & \textcolor{gelbbraun}{0} & \textcolor{gelbbraun}{0} & \textcolor{gelbbraun}{0} & \cellcolor{yellow}1 & \textcolor{gelbbraun}{0} & \cellcolor{yellow}1 & \cellcolor{yellow}1 & \textcolor{gelbbraun}{0} \\
14 & 14a & \textcolor{gelbbraun}{0} & \textcolor{gelbbraun}{0} & \textcolor{gelbbraun}{0} & \textcolor{gelbbraun}{0} & \textcolor{gelbbraun}{0} & \textcolor{gelbbraun}{0} & \textcolor{gelbbraun}{0} & \textcolor{gelbbraun}{0} & \textcolor{gelbbraun}{0} & \textcolor{gelbbraun}{0} & \textcolor{gelbbraun}{0} & \textcolor{gelbbraun}{0} & \textcolor{gelbbraun}{0} & \textcolor{gelbbraun}{0} & \textcolor{gelbbraun}{0} & \textcolor{gelbbraun}{0} & \textcolor{gelbbraun}{0} & \textcolor{gelbbraun}{0} & \textcolor{gelbbraun}{0} & \textcolor{gelbbraun}{0} & \textcolor{gelbbraun}{0} & \textcolor{gelbbraun}{0} & \textcolor{gelbbraun}{0} & \textcolor{gelbbraun}{0} & \textcolor{gelbbraun}{0} & \textcolor{gelbbraun}{0} & \textcolor{gelbbraun}{0} & \textcolor{gelbbraun}{0} & \textcolor{gelbbraun}{0} & \textcolor{gelbbraun}{0} \\
10 & 15a & \textcolor{gelbbraun}{0} & \textcolor{gelbbraun}{0} & \textcolor{gelbbraun}{0} & \textcolor{gelbbraun}{0} & \textcolor{gelbbraun}{0} & \textcolor{gelbbraun}{0} & \textcolor{gelbbraun}{0} & \cellcolor{yellow}1 & \textcolor{gelbbraun}{0} & \cellcolor{orange}2 & \textcolor{gelbbraun}{0} & \cellcolor{orange}2 & \textcolor{gelbbraun}{0} & \textcolor{gelbbraun}{0} & \textcolor{gelbbraun}{0} & \textcolor{gelbbraun}{0} & \textcolor{gelbbraun}{0} & \textcolor{gelbbraun}{0} & \textcolor{gelbbraun}{0} & \textcolor{gelbbraun}{0} & \textcolor{gelbbraun}{0} & \textcolor{gelbbraun}{0} & \textcolor{gelbbraun}{0} & \textcolor{gelbbraun}{0} & \textcolor{gelbbraun}{0} & \cellcolor{orange}2 & \textcolor{gelbbraun}{0} & \textcolor{gelbbraun}{0} & \textcolor{gelbbraun}{0} & \textcolor{gelbbraun}{0} \\
7 & 15b & \textcolor{gelbbraun}{0} & \textcolor{gelbbraun}{0} & \textcolor{gelbbraun}{0} & \textcolor{gelbbraun}{0} & \textcolor{gelbbraun}{0} & \textcolor{gelbbraun}{0} & \cellcolor{yellow}1 & \textcolor{gelbbraun}{0} & \cellcolor{orange}2 & \cellcolor{yellow}1 & \textcolor{gelbbraun}{0} & \cellcolor{yellow}1 & \textcolor{gelbbraun}{0} & \textcolor{gelbbraun}{0} & \textcolor{gelbbraun}{0} & \textcolor{gelbbraun}{0} & \textcolor{gelbbraun}{0} & \textcolor{gelbbraun}{0} & \textcolor{gelbbraun}{0} & \textcolor{gelbbraun}{0} & \textcolor{gelbbraun}{0} & \textcolor{gelbbraun}{0} & \textcolor{gelbbraun}{0} & \textcolor{gelbbraun}{0} & \textcolor{gelbbraun}{0} & \cellcolor{yellow}1 & \textcolor{gelbbraun}{0} & \textcolor{gelbbraun}{0} & \textcolor{gelbbraun}{0} & \textcolor{gelbbraun}{0} \\
11 & 16a & \textcolor{gelbbraun}{0} & \textcolor{gelbbraun}{0} & \textcolor{gelbbraun}{0} & \textcolor{gelbbraun}{0} & \textcolor{gelbbraun}{0} & \textcolor{gelbbraun}{0} & \textcolor{gelbbraun}{0} & \cellcolor{yellow}1 & \textcolor{gelbbraun}{0} & \textcolor{gelbbraun}{0} & \textcolor{gelbbraun}{0} & \textcolor{gelbbraun}{0} & \textcolor{gelbbraun}{0} & \textcolor{gelbbraun}{0} & \textcolor{gelbbraun}{0} & \textcolor{gelbbraun}{0} & \textcolor{gelbbraun}{0} & \textcolor{gelbbraun}{0} & \textcolor{gelbbraun}{0} & \textcolor{gelbbraun}{0} & \textcolor{gelbbraun}{0} & \textcolor{gelbbraun}{0} & \textcolor{gelbbraun}{0} & \textcolor{gelbbraun}{0} & \textcolor{gelbbraun}{0} & \cellcolor{orange}2 & \textcolor{gelbbraun}{0} & \textcolor{gelbbraun}{0} & \textcolor{gelbbraun}{0} & \textcolor{gelbbraun}{0} \\
4 & 16c & \textcolor{gelbbraun}{0} & \cellcolor{yellow}1 & \cellcolor{yellow}1 & \textcolor{gelbbraun}{0} & \cellcolor{red}3 & \cellcolor{orange}2 & \textcolor{gelbbraun}{0} & \cellcolor{yellow}1 & \cellcolor{orange}2 & \cellcolor{orange}2 & \textcolor{gelbbraun}{0} & \cellcolor{orange}2 & \textcolor{gelbbraun}{0} & \textcolor{gelbbraun}{0} & \textcolor{gelbbraun}{0} & \cellcolor{yellow}1 & \textcolor{gelbbraun}{0} & \cellcolor{yellow}1 & \textcolor{gelbbraun}{0} & \cellcolor{red}3 & \textcolor{gelbbraun}{0} & \textcolor{gelbbraun}{0} & \textcolor{gelbbraun}{0} & \cellcolor{yellow}1 & \textcolor{gelbbraun}{0} & \cellcolor{red}3 & \cellcolor{yellow}1 & \cellcolor{orange}2 & \textcolor{gelbbraun}{0} & \textcolor{gelbbraun}{0} \\
 & & 1b & 1c & 1d & 2b & 2c & 3b & 3e & 4e & 5c & 6c & 7a & 7d & 8b & 8e & 9a & 9c & 10a & 10c & 11b & 11d & 11e & 12m & 13b & 13c & 13d & 14a & 15a & 15b & 16a & 16c \\
\end{tabular}
}\\[1em]

\bibliography{./surv}

\end{document}